\documentclass[
 reprint,
superscriptaddress,
reprint,
preprintnumbers,
nofootinbib,
 amsmath,amssymb,
 aps,
]{revtex4-2}

\usepackage{caption}
\usepackage{subcaption}

\usepackage{float}
\usepackage{braket, amsmath} 
\usepackage{cancel}
\usepackage{bbm}
\usepackage{graphicx}
\usepackage{dcolumn}
\usepackage{bm}
\usepackage[colorlinks=true, linkcolor=black, citecolor=black, urlcolor=blue]{hyperref}

\usepackage[utf8]{inputenc}

\usepackage{tikz} 

\usepackage{subfiles}

\begin{document}

\title{Evolution of tripartite entanglement in three-qubit Quantum Gravity-Induced Entanglement of Masses (QGEM) with quantum decoherence}

\author{Pablo Guillermo Carmona Rufo}
\affiliation{
Instituto de Física Teórica, UAM-CSIC, C/ Nicolás Cabrera 13-15, Campus de Cantoblanco, 28049 Madrid, Spain.} 

\author{Anupam Mazumdar}
\affiliation{Van Swinderen Institute for Particle Physics and Gravity, University of Groningen, 9747AG, Groningen, The Netherlands.}

\author{Carlos Sabín}
\affiliation{Departamento de Física Teórica and CIAFF, Universidad Autónoma de Madrid, 28049, Madrid, Spain.}

\begin{abstract}
The recently introduced quantum gravity-induced entanglement of masses (QGEM) protocol aims to test the quantum nature of gravity by witnessing the entanglement produced by the virtual exchange of a graviton between two spatially superposed masses. Shortly after the original proposal, further improvements upon the experiment were suggested, including the addition of a third mass, showing that three-qubit setups can be more resilient to higher rates of decoherence caused by the interaction of the system with the environment. In this work, we investigate the type of tripartite entanglement generated in these three-qubit QGEM experiments when considering the effects of decoherence. We show that the gravitational interaction between the qubits is able to generate genuine tripartite entanglement between them, studying the corresponding parameter spaces and comparing the performance of the possible experimental configurations of the three qubits at allowing for the detection of genuine entanglement via an entanglement witness.
\end{abstract}
\maketitle
\section{Introduction} \label{one}

The consistent implementation of the gravitational interaction into the quantum framework is considered to be one of the outstanding problems in current physics. A new approach to the problem of quantum gravity has emerged in the past few years, focusing on proving the quantum nature of gravity without disclosing the underlying full theory. However, due to the extreme weakness of the gravitational interaction compared to the other fundamental forces of nature, its quantum effects are hard to detect. 

Recently, a proposal has emerged \cite{spin} to prove the quantum nature of gravity by witnessing the entanglement that it can produce between masses; see \cite{ICTS} as well. Proving that the detection of entanglement caused by a gravitational interaction shows the quantum nature of gravity relies on the Local Operations and Classical Communication (LOCC) principle \cite{correction,locality,mechanism}, which states that two quantum states cannot get entangled via a classical channel if they were not entangled beforehand. Therefore, if spacetime is of quantum nature, the two geometries with massive superpositions must be entangled by a quanta, known as the spin-2 graviton \cite{locality,mechanism,Danielson:2021egj,Carney:2018ofe,Carney:2022pku,Biswas:2022qto,Christodoulou:2022mkf,Vinckers:2023grv,Chakraborty:2023kel,warpedextra,Christodoulou:2018cmk,Rufo:2024ulr}.

The protocol introduced in \cite{spin} goes by the name of quantum gravity-induced entanglement of masses (QGEM) and is based on creating a macroscopic quantum superposition with two nanoparticles of mass $m\sim10^{-14} - 10^{-15}$ kg in an inhomogeneous magnetic field. The QGEM experiment can be used to probe several physical theories, such as the quantum weak equivalence principle \cite{equivalence}, testing beyond the Standard Model physics \cite{axions1,axions2} or constraining theories with massive gravity \cite{warpedextra}. Simultaneously to this work, another paper \cite{marletto} was released, where the authors proposed to witness the quantum nature of gravity by detecting entanglement. However, the detailed analysis of the graviton as a mediator, as well as some of the feasibility aspects of the experiment including the introduction of decoherence, were first discussed in \cite{spin}. The proposal in this work was met with extensive interest by the community, with plenty of extensions and variations suggested \cite{Belenchia:2018szb,Pedernales:2020nmf,nanoobject,plate1,loss,Chevalier:2020uvv,Toros:2020dbf,Christodoulou:2018cmk,Christodoulou:2018xiy,stern,Anupam3,Carney:2021yfw,Miki,Matsumura:2020law,Qvarfort:2018uag,Miao:2019pxw,Weiss:2021wdu,Cosco:2021,Datta:2021ywm,Krisnanda:2019glc,Haine:2018bwu}.

In \cite{Anupam1}, the authors explore a new QGEM design with three masses instead of two and show that this type of three-qubit protocol can perform better at generating entanglement, particularly when decoherence is taken into account, which comes at the cost of requiring more measurements to characterize entanglement with a good level of certainty. In that work, the authors fix the physical parameters of the system by considering a superposition width of $l=250$ $\mu$m and a minimum separation between any two particles of $d_\text{min}=200$ $\mu$m, kept constant for all setups. This minimal distance was proposed in \cite{Anupam1} so that the gravitational interaction between the qubits dominates over the Casimir-Polder interaction \cite{casimir}, which can pollute the experiment due to photon-mediated entanglement. Nonetheless, in \cite{Anupam2}, the authors mention that since the original proposal of three-qubit QGEM, the constraints for the physical parameters have greatly improved, with the newest experimental developments, including the introduction of a (super)conducting plate \cite{plate1,plate2} as well as diamagnetic trapping; see \cite{warpedextra,micrometer} for further detail. This has allowed for the parameters to reach the values of $l\geq10$ $\mu$m and $d_\text{min}=35$ $\mu$m for masses of $m\sim10^{-14}-10^{-15}$ kg. \footnote{At all times, we will make the assumption that the superposition width $l$ is time-independent. In practice, the time dependence of $l$ will be determined by the particular Stern-Gerlach protocol \cite{Anupam2}. }

In \cite{LIU2024129273}, this kind of setup is also studied, with the goal of analyzing the multipartite entanglement generated by the gravitational interaction and classifying the quantum states generated in the QGEM protocol according to the entanglement they present. In this paper, we go further than these two previous proposals, focusing on the genuine tripartite entanglement that is generated in the three-qubit QGEM setup both before and after the introduction of decoherence in the system. The main contribution of this work lies in the detection of genuine tripartite entanglement in the decohered QGEM states generated by three different experimental configurations through the usage of an entanglement witness, which also allows us to check the robustness to decoherence of the proposed setups when attempting to measure said witness. We find the optimal experimental parameters to detect genuine tripartite entanglement in the presence of decoherence and discuss their feasibility in current experimental setups.

\section{Three-qubit qgem setup}
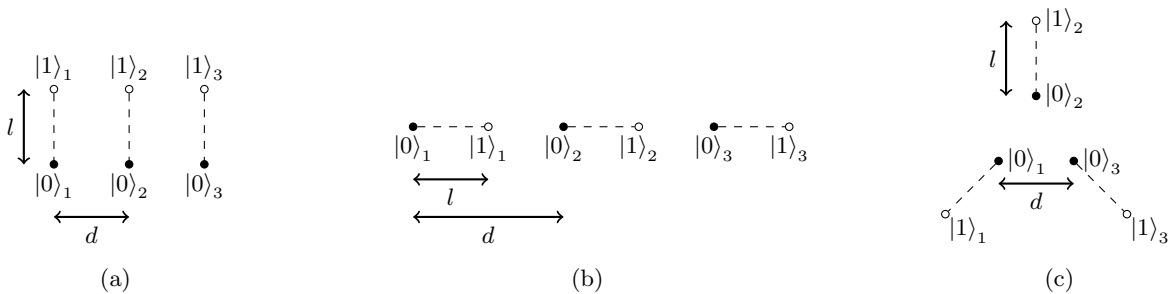
\begin{figure*}[t]
    \centering

    \begin{subfigure}{0.3\textwidth}
        \centering
        \begin{tikzpicture}
\node[circle, draw=black, fill=white, inner sep=1pt] (up1) at (-1,1) {};
\node[circle, draw=black, fill=black, inner sep=1pt] (down1) at (-1,0) {};
\node[circle, draw=black, fill=white, inner sep=1pt] (up2) at (0,1) {};
\node[circle, draw=black, fill=black, inner sep=1pt] (down2) at (0,0) {};
\node[circle, draw=black, fill=white, inner sep=1pt] (up3) at (1,1) {};
\node[circle, draw=black, fill=black, inner sep=1pt] (down3) at (1,0) {};

\draw[dashed] (up1) -- (down1);
\draw[dashed] (up2) -- (down2);
\draw[dashed] (up3) -- (down3);

\node[above=0pt] at (up1) {$\ket{1}_1$};
\node[below=0pt] at (down1) {$\ket{0}_1$};
\node[above=0pt] at (up2) {$\ket{1}_2$};
\node[below=0pt] at (down2) {$\ket{0}_2$};
\node[above=0pt] at (up3) {$\ket{1}_3$};
\node[below=0pt] at (down3) {$\ket{0}_3$};

\draw[<->, thick] (-1.4,0) -- (-1.4,1) node[midway,left] {$l$};
\draw[<->, thick] (-1,-0.7) -- (0,-0.7) node[midway,below] {$d$};

        \end{tikzpicture}
        \caption{}
    \end{subfigure}
    \hfill
    \begin{subfigure}{0.3\textwidth}
        \centering
        \begin{tikzpicture}
\node[circle, draw=black, fill=black, inner sep=1pt] (a) at (-2.5,-0.3) {};
\node[circle, draw=black, fill=white, inner sep=1pt] (b) at (-1.5,-0.3) {};
\node[circle, draw=black, fill=black, inner sep=1pt] (c) at (-0.5,-0.3) {};
\node[circle, draw=black, fill=white, inner sep=1pt] (d) at (0.5,-0.3) {};
\node[circle, draw=black, fill=black, inner sep=1pt] (e) at (1.5,-0.3) {};
\node[circle, draw=black, fill=white, inner sep=1pt] (f) at (2.5,-0.3) {};

\draw[dashed] (a) -- (b);
\draw[dashed] (c) -- (d);
\draw[dashed] (e) -- (f);

\node[below=0pt] at (a) {$\ket{0}_1$};
\node[below=0pt] at (b) {$\ket{1}_1$};
\node[below=0pt] at (c) {$\ket{0}_2$};
\node[below=0pt] at (d) {$\ket{1}_2$};
\node[below=0pt] at (e) {$\ket{0}_3$};
\node[below=0pt] at (f) {$\ket{1}_3$};

\draw[<->, thick] (-2.5,-1) -- (-1.5,-1) node[midway,below] {$l$};
\draw[<->, thick] (-2.5,-1.5) -- (-0.5,-1.5) node[midway,below] {$d$};

        \end{tikzpicture}
        \caption{}
    \end{subfigure}
    \hfill
    \begin{subfigure}{0.3\textwidth}
        \centering
        \begin{tikzpicture}
\node[circle, draw=black, fill=white, inner sep=1pt] (a) at (0,0.866025+1+0.7) {};
\node[circle, draw=black, fill=black, inner sep=1pt] (b) at (-0.5,0+0.7) {};
\node[circle, draw=black, fill=black, inner sep=1pt] (c) at (0.5,0+0.7) {};
\node[circle, draw=black, fill=black, inner sep=1pt] (d) at (0,0.866025+0.7) {};
\node[circle, draw=black, fill=white, inner sep=1pt] (e) at (-0.5-0.7071,-0.7071+0.7) {};
\node[circle, draw=black, fill=white, inner sep=1pt] (f) at (0.5+0.7071,-0.7071+0.7) {};

\draw[dashed] (a) -- (d);
\draw[dashed] (b) -- (e);
\draw[dashed] (c) -- (f);

\node[right=0pt] at (a) {$\ket{1}_2$};
\node[right=0pt] at (b) {$\ket{0}_1$};
\node[right=0pt] at (c) {$\ket{0}_3$};
\node[right=0pt] at (d) {$\ket{0}_2$};
\node[below right=-2pt] at (e) {$\ket{1}_1$};
\node[below right=-2pt] at (f) {$\ket{1}_3$};

\draw[<->, thick] (-0.4,0.866025+0.7) -- (-0.4,0.866025+1.7) node[midway,left] {$l$};
\draw[<->, thick] (-0.5,0.4) -- (0.5,0.4) node[midway,below] {$d$};

        \end{tikzpicture}
        \caption{}
    \end{subfigure}

    \caption{Experimental setups of the three-qubit QGEM experiment proposed in \cite{Anupam1,Anupam2}, which go by the name of (a) parallel, (b) linear and (c) star configurations. The distance $d$ is the distance between any two neighbouring $\ket{0}$ states, while $l$ is the superposition width of each qubit.}
    \label{fig:QGEM_setups}
\end{figure*}

The most general initial three-qubit unentangled state we can write is:
\begin{equation}
    \ket{\Psi_0}=\frac{1}{2\sqrt{2}}\bigotimes_{i=1}^3\left(\ket{0}_i+\ket{1}_i\right).
\label{eq:initial_state}
\end{equation}
Due to the gravitational interaction between them, each superposition state will pick up a phase given by $\phi=U\tau/\hbar$, where $U$ is the gravitational potential. The largest contribution from this potential comes from the tree-level exchange of a virtual graviton between the test masses \cite{micrometer}. In the non-relativistic regime of a perturbative quantum gravity theory in the weak field limit, for two particles of masses $m_1$ and $m_2$ separated by a distance $\hat{r}$, this gives the operator-valued potential $\hat{U}=Gm_1m_2/\hat{r}$, as argued in \cite{spin,locality,mechanism}, where $G$ represents the Newtonian gravitational constant. The total gravitational potential energy will then be $\hat{V}_{\text{tot}}=\hat{V}_{12}+\hat{V}_{23}+\hat{V}_{13}$, where each interaction between subsystems $i$ and $j$ will be:
\begin{equation}
  \hat{V}_{ij}=\frac{Gm_im_j}{|\hat{x}_i-\hat{x}_j|}.
\end{equation}
Therefore, using the notation where $j_i=0,1$ denotes the state of the $i^{\text{th}}$ qubit, we can write the evolved state once the qubits have undergone the gravitational interaction between them as:
\begin{equation}
    \ket{\Psi(t=\tau)}=\frac{1}{2\sqrt{2}}\sum_{j_1,j_2,j_3=0,1}e^{i\phi_{j_1j_2j_3}}\ket{j_1j_2j_3},
\label{eq:evolved_state}
\end{equation}
where $\phi_{j_1j_2j_3}$ are the phases picked up due to this gravitational interaction. Since these are determined by the distance between each pair of particles $|x_i-x_j|$, they will depend on the way the experimental setup is arranged. In this work, we will consider three different setups, denoted as parallel, linear and star configurations, which are displayed in Figure \ref{fig:QGEM_setups}. 

For the rest of this work, we will assume all interactions occur between particles of equal mass $m$. For the parallel case, we will have:
\begin{equation}
    \phi_{j_1 j_2 j_3}^{(\parallel)} = \frac{1}{\hbar} \sum_{\substack{i,k=1,2,3 \\ i<k}} \frac{Gm^2\tau}{\sqrt{\left(d\left(k-i\right)\right)^2 + (l(j_i - j_k))^2}}.
\end{equation}
For the linear setup:
\begin{equation}
    \phi_{j_1 j_2 j_3}^{(-)} = \frac{1}{\hbar} \sum_{\substack{i,k=1,2,3 \\ i<k}} \frac{Gm^2\tau}{(k - i)d + l(j_k - j_i)},
\end{equation}
while for the star configuration, the phases will be:
\begin{equation}
    \phi_{j_1 j_2 j_3}^{(*)} = \frac{1}{\hbar} \sum_{\substack{i,k=1,2,3 \\ i<k}} \frac{Gm^2\tau}{r_{ik}},
\end{equation}
where the distance between two superposition instances $i$ and $k$ is:
\begin{equation}
    r_{ik}=(1-j_ij_k)d+j_ij_k(d+\sqrt{3}l)+\\|j_k-j_i|R,
\end{equation}
with
\begin{equation}
    R=\sqrt{d^2+\sqrt{3}d l+l^2}-d.
\end{equation}

 Considering what is detailed in Section \ref{one} about the most recent experimental improvements in three-qubit QGEM setups, the values we will use in certain cases through this work when needed will be $m=10^{-14}$ kg, $d_\text{min}=35$ $\mu$m and $l=10$ $\mu$m, which are consistent with the feasible range reported in \cite{spin,nanoobject}. Thus, we will have $d=d_\text{min}=35$ $\mu$m for the parallel and star setups but $d=d_\text{min}+l=45$ $\mu$m for the linear one. Plus, whenever it has to be fixed, the value for the interaction time is chosen to be $\tau=2.5$ s, which is both experimentally accessible while also large enough for entanglement to be detectable \cite{locality,stern}.
 
From (\ref{eq:evolved_state}), we can find that the density matrix of the system $\rho(\tau)$ is given by:
\begin{equation}
    \rho(\tau)=\frac{1}{8}\sum_{\substack{ j_1,j_2,j_3=0,1 \\ j'_1,j'_2,j'_3=0,1 }}e^{i(\phi_{j_1j_2j_3}-\phi_{j'_1j'_2j'_3})}\bigotimes_{i,i'=1,2,3}\ket{j_i}\bra{j'_{i'}}.
\end{equation}

\section{Decoherence}
So far, we have assumed that the system is not affected by its environment at all. Since this is not a very realistic approach, we are interested in recreating the interaction between the system and the environment, which makes the system share part of its information and thus potentially lose it. This loss of information is defined as decoherence, and, even though it is determined by the particular interactions with the environment, it can be generally modeled following \cite{Anupam1,schlosshauer2019quantumtoclassicaltransitiondecoherence}, assuming that the environmental state $\ket{E_i}$ couples to the system position state $\ket{\vec{x}(i)}$ and rewriting Equation (\ref{eq:initial_state}) in the following way:
\begin{equation}
    \ket{\Psi_0}=\frac{1}{2\sqrt{2}}\sum_{j_1,j_2,j_3=0,1}\ket{\vec{x}(j_1)\vec{x}(j_2)\vec{x}(j_3)}\ket{E_{j_1}E_{j_2}E_{j_3}}.
\end{equation}
We can extract the system's entanglement by tracing out the environmental degrees of freedom $\rho_s=\text{Tr}_e(\rho)=\sum_i\bra{E_i}\rho\ket{E_i}$:
\begin{multline}
    \rho_S(0)=\frac18\sum_{j_1,j_2,j_3=0,1}\ket{j_1j_2j_3}\bra{j_1j_2j_3}\\+\frac18\sum^{j_1j_2j_3\neq j'_1j'_2j'_3}_{\substack{ j_1,j_2,j_3=0,1 \\ j'_1,j'_2,j'_3=0,1 }}\ket{j_1j_2j_3}\bra{j'_1j'_2j'_3}\bra{E_{j_1}E_{j_2}E_{j_3}}E_{j'_1}E_{j'_2}E_{j'_3}\rangle.
\label{eq:decoh_1}
\end{multline}
As we see, the system loses coherence as $\bra{E_{j_1}E_{j_2}E_{j_3}}E_{j'_1}E_{j'_2}E_{j'_3}\rangle\rightarrow 0$. As done in plenty of decoherence models \cite{schlosshauer2019quantumtoclassicaltransitiondecoherence}, we will assume that this overlap between the environmental states decreases exponentially over time with a rate of $\gamma$, known as the decoherence rate:
\begin{equation}
\langle E_j(t)\ket{E_{j'}(t)}\propto e^{-\gamma t}~~~~~j\neq j' .
\end{equation}
Then, time evolving the state in (\ref{eq:decoh_1}):
\begin{multline}
    \rho_S(\tau)=\frac18\sum_{j_1,j_2,j_3=0,1}\ket{j_1j_2j_3}\bra{j_1j_2j_3}\\+\frac18\sum^{j_1j_2j_3\neq j'_1j'_2j'_3}_{\substack{ j_1,j_2,j_3=0,1 \\ j'_1,j'_2,j'_3=0,1 }}e^{-\delta\gamma\tau}e^{i(\phi_{j_1j_2j_3}-\phi_{j'_1j'_2j'_3})}\ket{j_1j_2j_3}\bra{j'_1j'_2j'_3},
\end{multline}
where $\delta=3-\delta_{j_1j'_1}-\delta_{j_2j'_2}-\delta_{j_3j'_3}$. With this exponential decay, the off-diagonal terms go to zero over time, which makes it impossible to measure the entanglement, since it progressively transforms the experimental quantum state into a maximally mixed state. A more detailed breakdown of the interaction with the environment is given in \cite{Anupam1}, following the analysis in \cite{vandekamp}.

\section{Tripartite entanglement measures and witness}

\subsection{Tripartite negativity}
In \cite{LIU2024129273}, the authors study the tripartite entanglement in the state generated by the three-qubit QGEM parallel system using the tripartite negativity, defined in \cite{Sabin2008} as:
\begin{equation}
    \mathcal{N}_{ABC}(\rho)=\left(\mathcal{N}_{A-BC}\mathcal{N}_{B-AC}\mathcal{N}_{C-AB}\right)^{\frac13},
\label{eq:trip_neg_def}
\end{equation}
where the negativity of each bipartition is given by:
\begin{equation}
N_{I-JK}=-2\sum_i\sigma_i\left(\rho^{TI}\right),
\label{eq:neg_def}
\end{equation}
with $\sigma_i\left(\rho^{TI}\right)$ being the negative eigenvalues of $\rho^{TI}$, the partial transpose of $\rho$ with respect to subsystem $I$. Therefore, $\mathcal{N}>0$ is a sufficient condition for tripartite entanglement in the system. Nonetheless, this is only studied in \cite{LIU2024129273} with the evolved state remaining pure, with no introduction of decoherence. In \cite{Anupam1,Anupam2}, decoherence is introduced in the system in the same way as in this work, but the entanglement is studied through the usage of the PPT witness, which can only examine the entanglement of one partition with the rest of the system. This however does not guarantee the presence of tripartite entanglement, which is what we will focus on here.

\subsection{Three-tangle}
Another three-mode entanglement measure we make use of is the three-tangle $\tau_{\mathcal{A}\mathcal{B}\mathcal{C}}$ \cite{threetangle,distributed}, defined as:
\begin{equation}
    \tau_{\mathcal{A}\mathcal{B}\mathcal{C}}=\tau_{\mathcal{A}|\mathcal{B}\mathcal{C}}-\tau_{\mathcal{A}|\mathcal{B}}-\tau_{\mathcal{A}|\mathcal{C}}.
\end{equation}
There is more than one way to calculate this measure for pure three-qubit states, but for this work, we will use:
\begin{equation}
    \tau_{\mathcal{A}\mathcal{B}\mathcal{C}}=4|d_1-2d_2+4d_3|,
\label{eq:tangle}
\end{equation}
where the $d_i$ are given by the following relations between the coefficients of any given three-qubit state:
\begin{equation}    
\begin{aligned}
    d_1 =~ &a_{000}^2a_{111}^2+a_{001}^2a_{110}^2+a_{010}^2a_{101}^2+a_{100}^2a_{011}^2\\
    d_2 =~ &a_{000}a_{111}a_{011}a_{100}+a_{000}a_{111}a_{101}a_{010}\\
    +~& a_{000}a_{111}a_{110}a_{001}+a_{011}a_{100}a_{101}a_{010}\\
    +~& a_{011}a_{100}a_{110}a_{001}+a_{101}a_{010}a_{110}a_{001}\\
    d_3 =~ &a_{000}a_{110}a_{101}a_{011}+a_{111}a_{001}a_{010}a_{100}.
\end{aligned}
\end{equation}

\subsection{Genuine tripartite entanglement witness}

Entanglement measures such as the tripartite negativity or the three-tangle allow us to detect the presence of some form of tripartite entanglement in the considered system, namely that the state is neither separable nor biseparable with respect to any particular bipartition. However, there is still the possibility that the state could be written as a sum of biseparable states each belonging to a different bipartition, which would mean that the state is not genuinely entangled. To rule out this possibility we need to resort to  observables called entanglement witnesses $\mathcal{W}$, which can confirm the presence of entanglement in a state $\rho$ whenever $\text{Tr}(\mathcal{W}\rho)<0$ \cite{witness3}. The idea behind constructing a genuine multipartite entanglement witness is based on creating a witness around a genuinely entangled state $\ket{\psi}$ that is able to reveal genuine entanglement in a vicinity of that state. For three-qubit states, one makes the ansatz \cite{witness3}:
\begin{equation}
    \mathcal{W}=\chi\mathbbm{1}-\ket{\psi}\bra{\psi},
\label{eq:witness_def}
\end{equation}
where $\chi$ is defined as the maximal overlap between $\ket{\psi}$ and the pure biseparable states. This computation is straightforward \cite{alpha}, since any pure biseparable state is separable with respect to some bipartition, and for this case we know that this overlap is given by the maximal squared Schmidt coefficient, so in order to construct an optimal witness one can take:
\begin{equation}
    \chi=\underset{\text{bipartitions}~bp}{\text{max}}\left\{\underset{\text{Schmidt coefficients}~s_k(bp)}{\text{max}}\{[s_k(bp)]^2\}\right\}.
\label{eq:chi}
\end{equation}

Our goal in this work is to study the evolution of the tripartite entanglement generated in each experimental setup when decoherence is introduced in the QGEM system. We will indicate the range of values of the decoherence parameter $\gamma$ for which both the negativity and the witness are able to detect their respective types of entanglement, as well as show the advantage of calculating the system's tripartite entanglement instead of just one bipartition, highlighting the difference between the two. Moreover, we will make use of the three-tangle to verify the classification of states generated by the parallel QGEM setup given in \cite{LIU2024129273}, as well as create a new one for the linear case.

\section{Results}

\subsection{Parallel setup}
In the parallel setup, some of the phases share a value due to the symmetry of the system. Particularly, there will only be three different phases, $\phi_{1}\equiv\phi_{000}=\phi_{111}$, $\phi_{2}\equiv\phi_{001}=\phi_{011}=\phi_{100}=\phi_{110}=$ and $\phi_{3}\equiv\phi_{010}=\phi_{010}$, which will be given by:
\begin{equation}
\begin{aligned}
\phi_1 &= \frac{5G m^2 \tau}{2 \hbar d}, \\
\phi_2 &= \frac{G m^2 \tau}{\hbar} \left( \frac{1}{d} + \frac{1}{\sqrt{4d^2 + l^2}} + \frac{1}{\sqrt{d^2 + l^2}} \right), \\
\phi_3 &= \frac{G m^2 \tau}{\hbar} \left( \frac{1}{2d} + \frac{2}{\sqrt{d^2 + l^2}} \right).
\end{aligned}
\end{equation}
Writing $\Delta\phi_i=\phi_i-\phi_1$, the time-evolved state before decoherence will be:
\begin{multline}
    \ket{\psi_{\text{par}}}=\frac{e^{i\phi_1}}{2\sqrt{2}}\Bigg[\ket{000}+\ket{111}+\\
    e^{i\Delta\phi_2}\big(\ket{001}+\ket{011}+\ket{100}+\ket{110}\big)\\
    +e^{i\Delta\phi_3}\big(\ket{010}+\ket{101}\big)\Bigg].
\end{multline}
Following \cite{LIU2024129273}, we know that this state can have three different forms depending on the values of the phases:
\begin{itemize}
    \item[(i)] When $\Delta\phi_3=2n\pi$, $n\in \mathbb{Z}$, the state will be separable. Moreover, it will be fully inseparable if $\Delta\phi_2=n\pi$.
    
    \item[(ii)] When $\Delta\phi_3=\left(2n+1\right)\pi$, $n\in \mathbb{Z}$, the state will become LU equivalent to a GHZ state and therefore genuinely entangled.
    
    \item[(iii)] When neither of the two previous conditions is fulfilled, the most general case will be when the state becomes what the authors in \cite{LIU2024129273} define as ``GHZ-type'' states, that is, states that are equivalent to a GHZ state under stochastic locar operations and classical communications (SLOCC).
\end{itemize}
If we calculate the three-tangle of the system using (\ref{eq:tangle}), setting the global phase $e^{i\phi_1}=1$ and defining $\alpha=e^{i\Delta\phi_2}$ and $\beta=e^{i\Delta\phi_3}$, we get:
\begin{equation}
\tau_{\text{par}}(\alpha,\beta)=\frac{1}{16}\Big|1 + \beta^4 - 2\beta^2 - 4\alpha^2 + 8\alpha^2\beta - 4\alpha^2\beta^2\Big|.
\end{equation}
This is in accordance with the aforementioned classification of generated states made in \cite{LIU2024129273}, since:
\begin{itemize}
    \item[(i)]  $\tau_{\text{par}}(\alpha,1)=0$, since the states are separable. It is worth remarking that this result for the value of the three-tangle agrees with the fact that the states are separable, as shown in \cite{LIU2024129273}, but does not directly prove it, since one can find states that are not separable and yet have a null three-tangle, such as the family of $W$ states $\ket{W}=a\ket{100}+b\ket{010}+c\ket{001}$.

    \item[(ii)] $\tau_{\text{par}}(\alpha,-1)=|\alpha^2|=1$, which is only fulfilled by states in the GHZ family.
    \item[(iii)] Any other state has $0<\tau_{\text{par}}(\alpha,\beta)<1$.
\end{itemize}

\begin{figure*}[t]
    \centering

    \begin{subfigure}{0.49\textwidth}
        \includegraphics[width=
\linewidth]{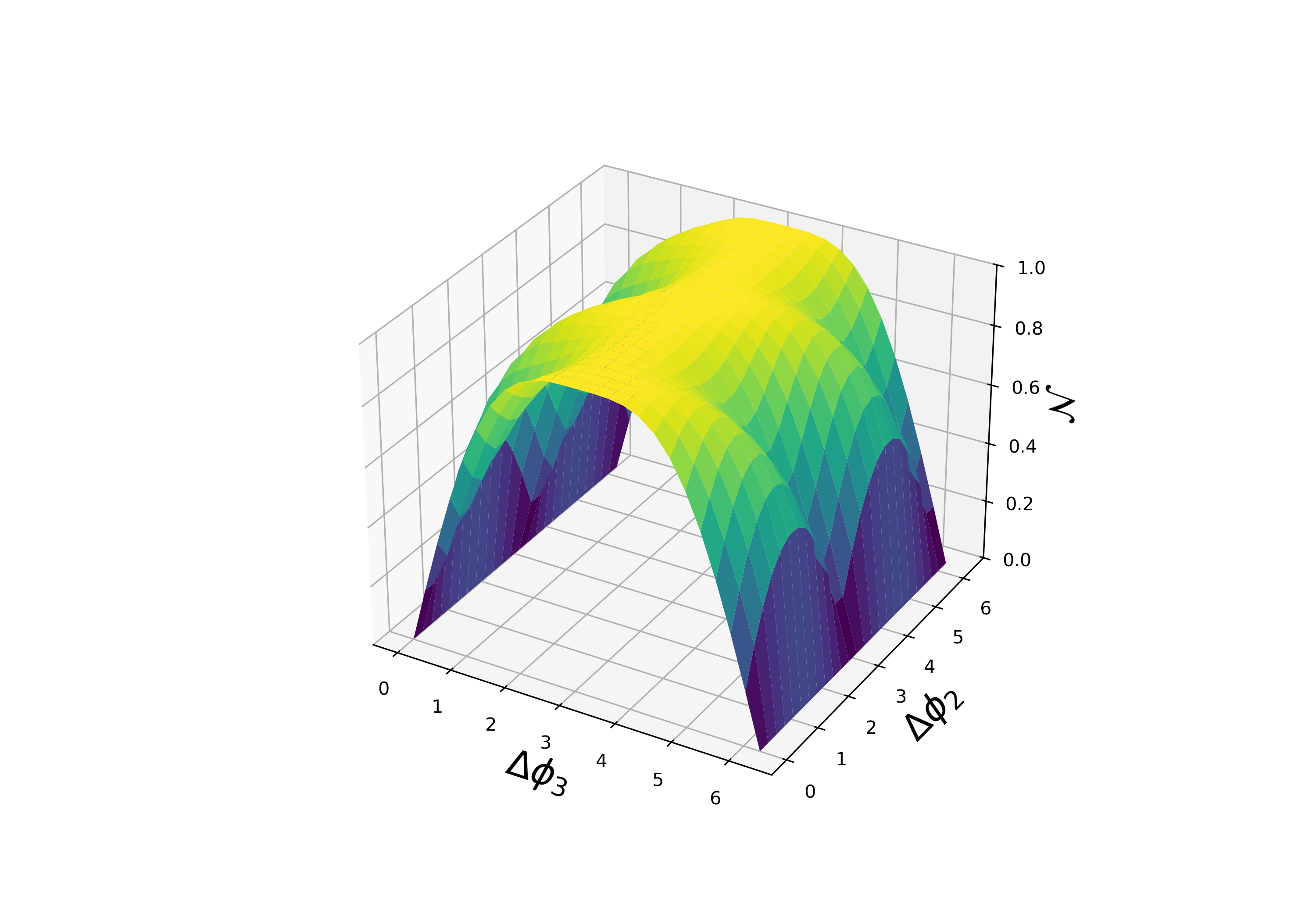}
        \centering
        \caption{}
    \end{subfigure}
    \hfill
    \begin{subfigure}{0.49\textwidth}
    \includegraphics[width=
\linewidth]{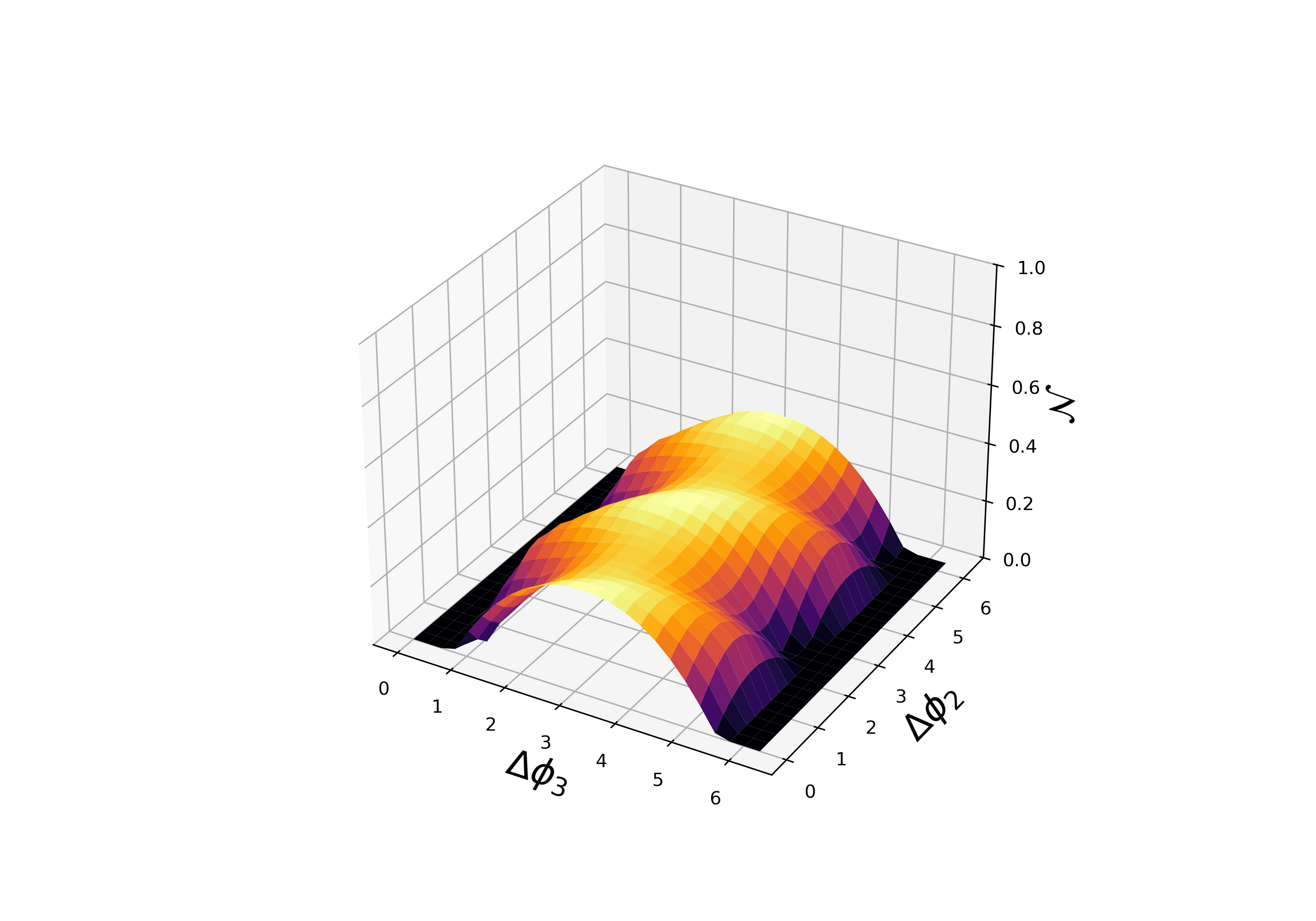}
        \centering

        \caption{}
    \end{subfigure}
    \caption{Tripartite negativity from Equation (\ref{eq:trip_neg_def}) in the parallel setup as a function of the phases $\Delta\phi_2$ and $\Delta\phi_3$, with (a) $\gamma=0$ Hz and (b) $\gamma=0.2$ Hz and $\tau=2.5$ s. Before the decoherence is introduced, the states with $\Delta\phi_3=2n\pi$ have no entanglement, while the GHZ states can be found in the $\Delta\phi_3=\left(2n+1\right)\pi$, as predicted. The negativity (and thus, the entanglement) gets smaller as the decoherence grows, up to $\gamma\simeq0.5$ Hz, where it is negligible. When decoherence is present, the maximum values for the tripartite negativity are found for $\Delta\phi_2=n\pi$ and $\Delta\phi_3=\left(2n+1\right)\pi$.}
    \label{fig:neg01par_3d}
\end{figure*}

We are also able to represent the tripartite negativity in the system as a function of the two phase differences, before the introduction of decoherence, as shown in Figure \ref{fig:neg01par_3d}. We can see that the presence of the decoherence decreases the amount of negativity detectable, up to the point where, for the chosen value of the experimental time $\tau$, makes it fully negligible in the whole phase space for $\gamma\simeq0.5$ Hz.

\begin{figure*}[t]
    \centering

    \begin{subfigure}{0.49\textwidth}
        \includegraphics[width=
\linewidth]{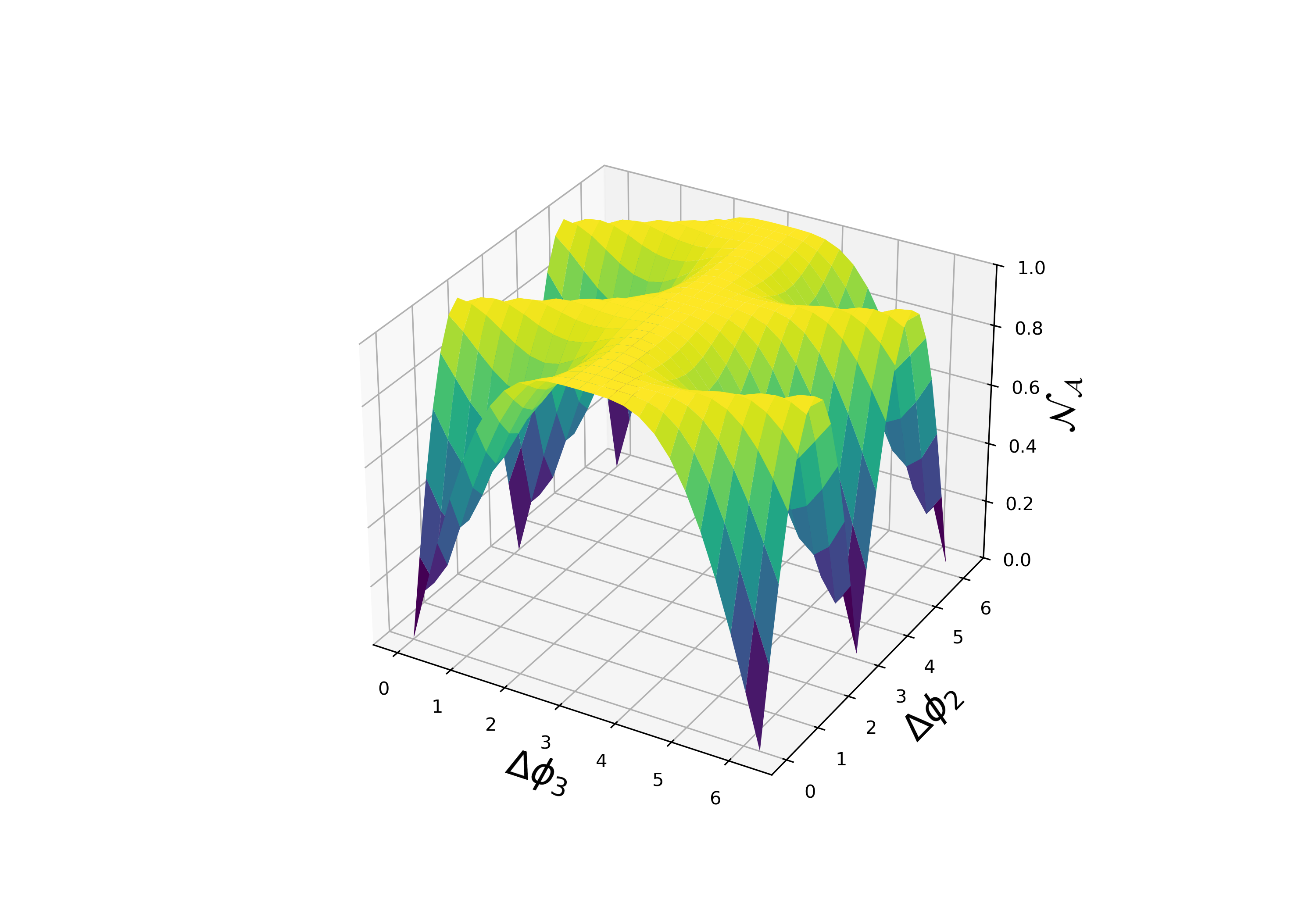}
        \centering
        \caption{}
    \end{subfigure}
    \hfill
    \begin{subfigure}{0.49\textwidth}
    \includegraphics[width=
\linewidth]{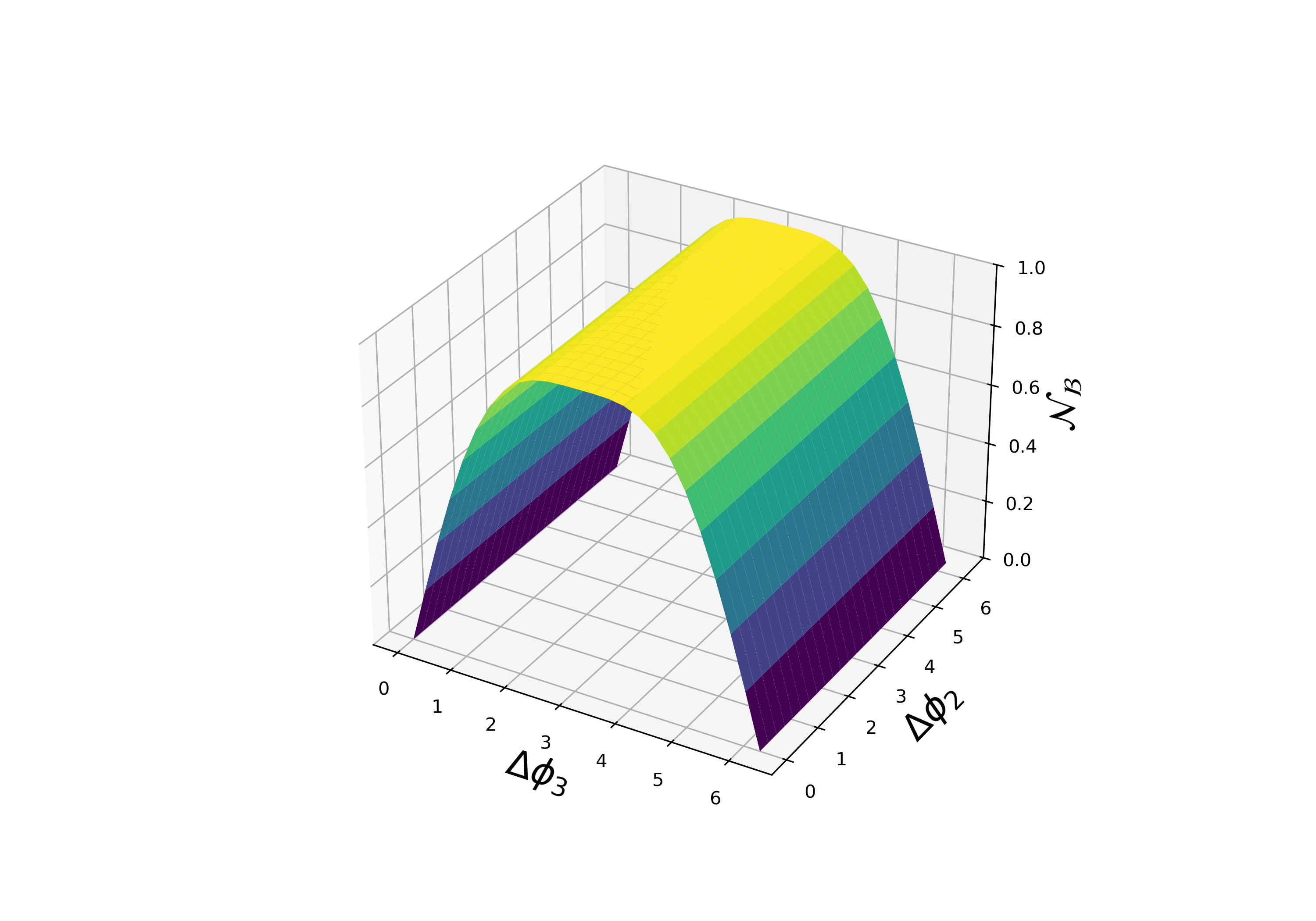}
        \centering

        \caption{}
    \end{subfigure}
    \caption{Negativity from Equation (\ref{eq:neg_def}) of the (a) first and (b) second bipartition in the parallel setup as a function of the phases $\Delta\phi_2$ and $\Delta\phi_3$, with $\gamma=0$ Hz. The mass $m$, separation $d_\text{min}$ and superposition width $l$ are not fixed but kept as free parameters.}
    \label{fig:Neg_bipar_static_AB_0}
\end{figure*}

\begin{figure*}[t]
    \centering

    \begin{subfigure}{0.49\textwidth}
        \includegraphics[width=
\linewidth]{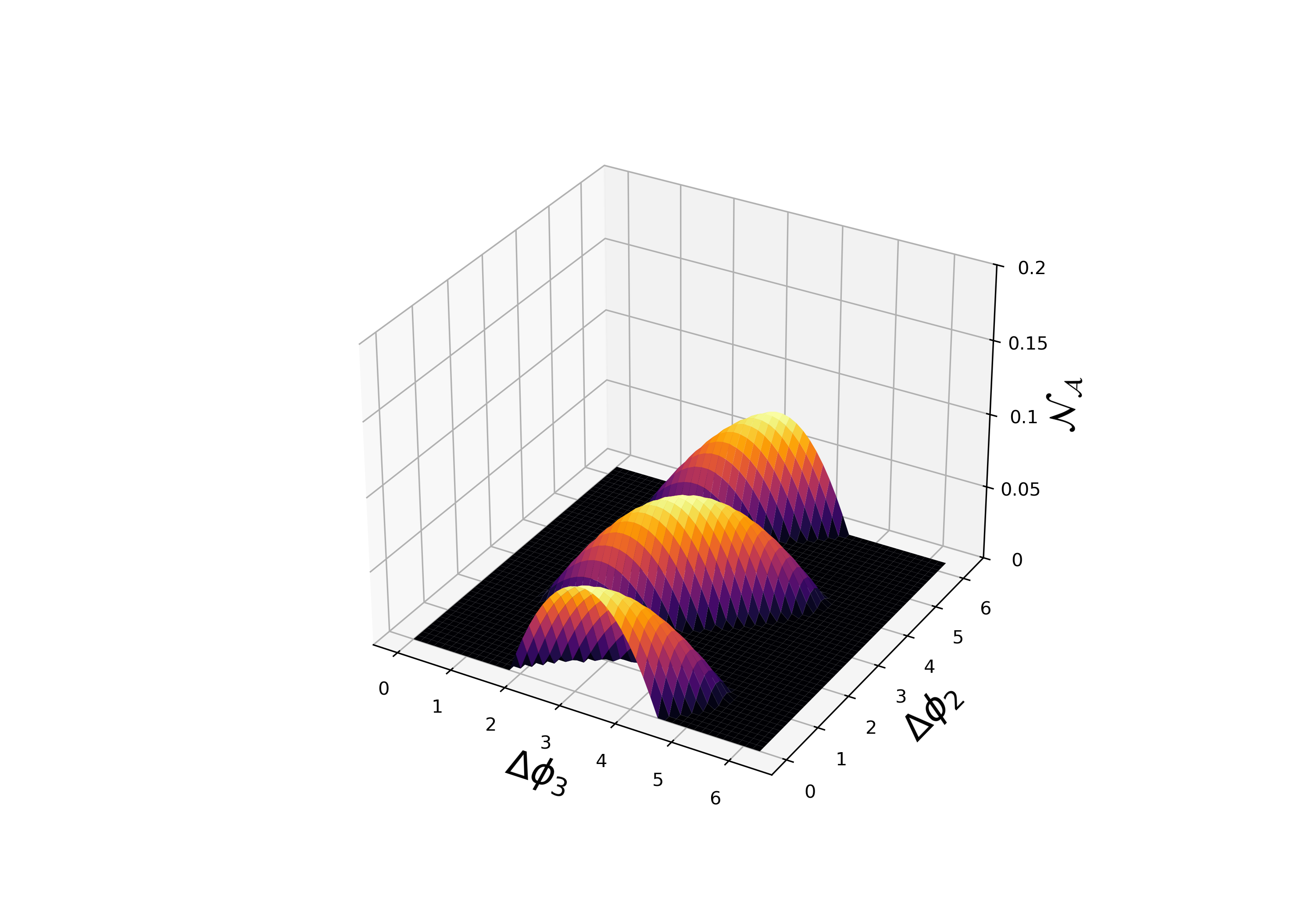}
        \centering
        \caption{}
    \end{subfigure}
    \hfill
    \begin{subfigure}{0.49\textwidth}
    \includegraphics[width=
\linewidth]{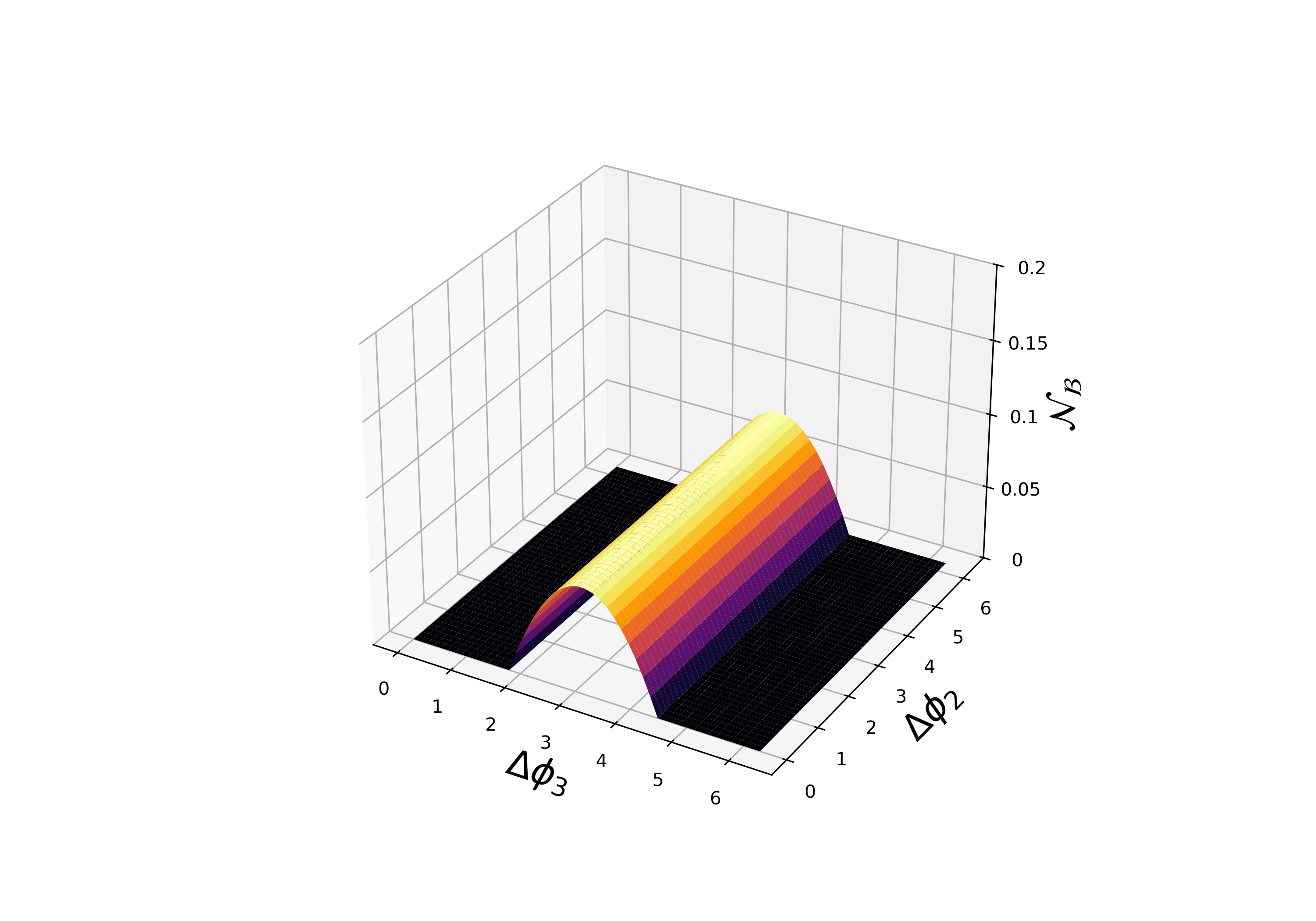}
        \centering

        \caption{}
    \end{subfigure}
    \caption{Negativity (\ref{eq:neg_def}) of the (a) first and (b) second bipartition in the parallel setup as a function of the phases $\Delta\phi_2$ and $\Delta\phi_3$, with $\gamma=0.4$ Hz and $\tau=2.5$ s. The mass $m$, separation $d_\text{min}$ and superposition width $l$ are not fixed but kept as free parameters.}
    \label{fig:Neg_bipar_static_AB_1}
\end{figure*}

The advantage of using the tripartite negativity (or the three-tangle) as our entanglement measure is that it allows us to claim the presence of tripartite entanglement in the system, as opposed to if we were to use the negativity or any other measure that, for three-qubit systems, only studies the entanglement of one bipartition, such as the entanglement entropy or the PPT witness, as done in \cite{Anupam1,Anupam2}. Looking at (\ref{eq:trip_neg_def}), we can see that the negativity of one of the bipartitions being null for any pair of values in the phase space is enough to determine that the tripartite negativity vanishes at that point. Thus, this means that we either need to examine each possible bipartition separately or make use of a tripartite entanglement measure if we wish to study the tripartite entanglement. In Figure \ref{fig:Neg_bipar_static_AB_0} we show the negativity of the bipartitions corresponding to the first two qubits, $\mathcal{N}_\mathcal{A}$ and $\mathcal{N}_\mathcal{B}$ (due to the symmetry in the setup, we will have $\mathcal{N}_\mathcal{A}=\mathcal{N}_\mathcal{C}$). 
The figures show that even though the entanglement distribution is different for each bipartition, there is not a point in the phase space where the negativity is null for the first one but is different from zero for the second. This hints at the possibility that studying the entanglement of the second bipartition might be enough to determine the values of the phases for which tripartite entanglement is present. However, this changes when decoherence is introduced, as shown in Figure \ref{fig:Neg_bipar_static_AB_1}: Around $\Delta\phi_3=\pi$, there are certain values of $\Delta\phi_2$ for which the entanglement does not vanish for the second bipartition but does for the first.

\begin{figure*}[t]
    \centering

    \begin{subfigure}{0.49\textwidth}
        \includegraphics[width=
\linewidth]{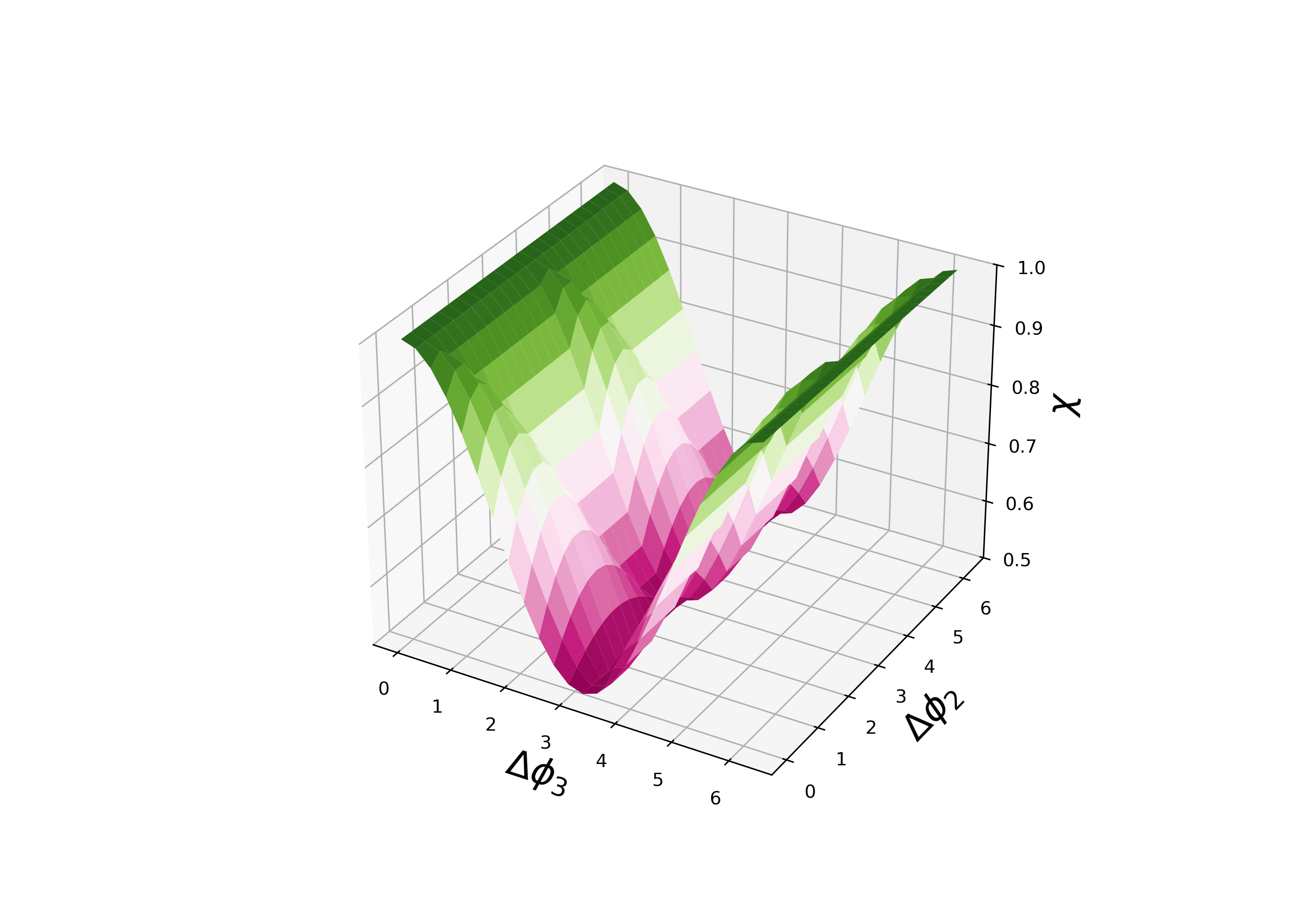}
        \centering
        \caption{}
    \end{subfigure}
    \hfill
    \begin{subfigure}{0.49\textwidth}
    \includegraphics[width=
\linewidth]{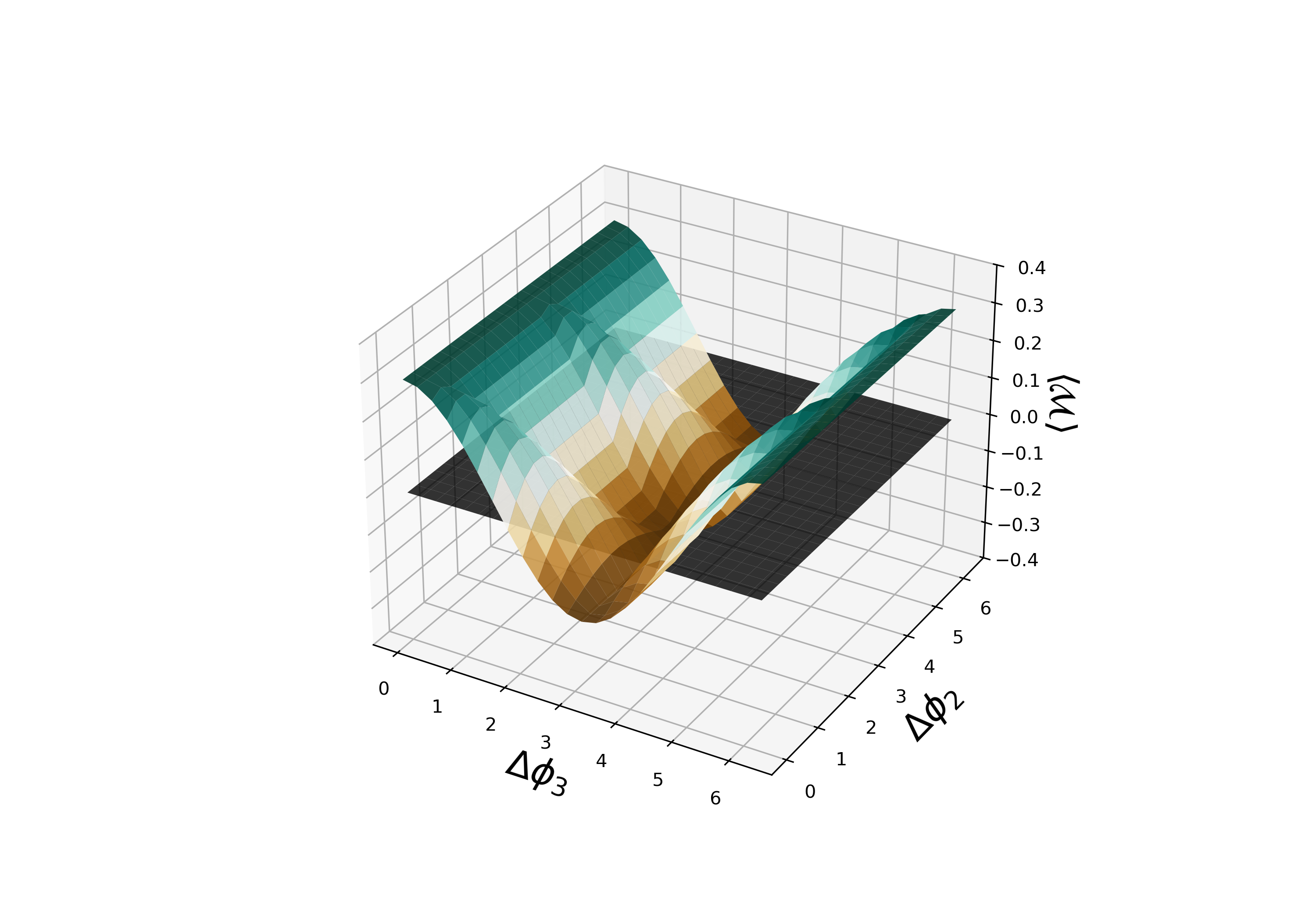}
        \centering

        \caption{}
    \end{subfigure}
    \caption{ (a) $\chi$ parameter from Equation (\ref{eq:chi}) necessary to calculate the (b) genuine tripartite entanglement witness from Equation (\ref{eq:witness_def}) in the parallel setup as a function of the phases $\Delta\phi_2$ and $\Delta\phi_3$, with $\gamma=0.1$ Hz and $\tau=2.5$ s. The parameter resides in the range $[0.5,1]$ for the whole phase space, with $\chi=1$ for the separable states and $\chi=0.5$ for the GHZ states. The mass $m$, separation $d_\text{min}$ and superposition width $l$ are not fixed but kept as free parameters.}
    \label{fig:witness}
\end{figure*}

After this, we can represent in a similar fashion the values of the genuine entanglement witness. For $\gamma=0$, we know that every state, except for the separable ones, has genuine tripartite entanglement, so we can construct our witness around them. Thus, every point in the phase space will lead to a different value of $\chi$ and therefore a different witness being used. As shown in Figure \ref{fig:witness}, for $\gamma=0.1$ s and $\tau=2.5$ s the genuine entanglement can still be detected at certain points, most notably around $\Delta\phi_3=\pi$, where the interaction produces GHZ states. For $\gamma\simeq0.2$ Hz, the whole surface falls above the $\langle\mathcal{W}\rangle=0$ plane and the ability to signal entanglement is lost.

\subsection{Linear setup}
In the linear setup, there are four different phases, $\varphi_{1}\equiv\varphi_{000}=\varphi_{111}$, $\varphi_{2}\equiv\varphi_{001}=\varphi_{011}$, $\varphi_{3}\equiv\varphi_{010}=\varphi_{010}$ and $\varphi_4=\varphi_{100}=\varphi_{110}$:
\begin{equation}
\begin{aligned}
\varphi_1 &= \frac{5G m^2 \tau}{2 \hbar d}, \\
\varphi_2 &= \frac{G m^2 \tau}{\hbar} \left( \frac{1}{d} + \frac{1}{d+l} + \frac{1}{2d+l} \right), \\
\varphi_3 &= \frac{G m^2 \tau}{\hbar} \left( \frac{1}{2d} + \frac{1}{d+l} + \frac{1}{d-l} \right), \\
\varphi_4 &= \frac{G m^2 \tau}{\hbar} \left( \frac{1}{d} + \frac{1}{d-l} + \frac{1}{2d-l} \right),
\end{aligned}
\end{equation}
Due to this, we will not be able to represent the entanglement measures in the same way as Figures \ref{fig:neg01par_3d}-\ref{fig:witness}, but there still is plenty of information we can extract. Using the same notation as in the previous case, $\Delta\varphi_i=\varphi_i-\varphi_1$:
\begin{multline}
    \ket{\psi_{\text{lin}}}=\frac{e^{i\varphi_1}}{2\sqrt{2}}\Bigg[\ket{000}+\ket{111}+\\
    e^{i\Delta\varphi_2}\big(\ket{001}+\ket{011}\big)+e^{i\Delta\varphi_3}\big(\ket{010}+\ket{101}\big)\\
    +e^{i\Delta\varphi_4}\big(\ket{100}+\ket{110}\big)\Bigg].
\label{eq:phi_lin}
\end{multline}
Inspecting this state, we recognize that its physical properties will be equivalent to those of the parallel case for $\varphi_2=\varphi_4$, which happens for $l=\sqrt{5/2}d$. For the cases in which $\varphi_2\neq\varphi_4$, we will use the three-tangle to study the entanglement. Similarly as before, we write $\alpha=e^{i\Delta\varphi_2}$, $\beta=e^{i\Delta\varphi_3}$ and $\lambda=e^{i\Delta\varphi_4}$ and make use of (\ref{eq:tangle}):
\begin{multline}
\tau_{\text{lin}}(\alpha,\beta,\lambda)\\
=\frac{1}{16}\Big|1+\beta^4-2\beta^2-4\alpha \lambda+8\alpha\beta\lambda-4\alpha\beta^2\lambda\Big|.
\end{multline}
This allows us to create a similar classification as the one shown in the previous section, by checking the three-tangle of the state for different values of the parameters. The relation $\tau_{\text{lin}}(\alpha,1,\lambda)=0$ still holds for $\alpha\neq\lambda$, which hints at the possibility that the state is also separable in this case (but does not prove it). Indeed, we can see that we can rewrite Equation (\ref{eq:phi_lin}) in the following form \cite{LIU2024129273}:
\begin{multline}
    \ket{\psi_{\text{lin}}}=\frac{1}{2\sqrt{2}}\Bigg[\ket{0}\Big(\ket{00}+\alpha\big(\ket{01}+\ket{11}\big)+\beta\ket{10}\Big)\\
    +\ket{1}\Big(\ket{11}+\lambda\big(\ket{00}+\ket{01}\big)+\beta\ket{01}\Big)\Bigg],
\end{multline}
which, for $\beta=1$, can be expressed as a separable state:
\begin{multline}
        \ket{\psi_{\text{lin}}}=\frac{1}{2\sqrt{2}}\Bigg[\ket{0}\otimes\left(\ket{0}+\ket{1}\right)\otimes\left(\ket{0}+\alpha\ket{1}\right)\\
    +\ket{1}\otimes\left(\ket{0}+\ket{1}\right)\otimes\left(\lambda\ket{0}+\ket{1}\right)\Bigg],
\end{multline}
which will be fully separable for $\alpha^{-1}=\lambda$ or $\varphi_3=-\varphi_2$ (in the parallel case, with $\alpha=\lambda$, this will hold for $\alpha=\lambda=\pm1$, as shown in \cite{LIU2024129273}). Furthermore, we can also see that $\tau_{\text{lin}}(\alpha,-1,\lambda)=|\alpha\lambda|=1$, which shows that these states are also GHZ and thus possess the same type of genuine entanglement.

\subsection{Star setup}
Just like in the linear configuration, there are four different phases in the star setup, $\Phi_1\equiv\Phi_{000}$, $\Phi_2\equiv\Phi_{111}$, $\Phi_3\equiv\Phi_{001}=\Phi_{010}=\Phi_{100}$ and $\Phi_4\equiv\Phi_{011}=\Phi_{101}=\Phi_{110}$
\begin{equation}
\begin{aligned}
\Phi_{1} &= \frac{3G m^2 \tau}{d\hbar}, \\
\Phi_{2} &= \frac{3G m^2 \tau}{\left (d+\sqrt{3}l\right)\hbar},\\
\Phi_{3} &= \frac{G m^2 \tau}{\hbar}\left(\frac{1}{d}+\frac{2}{d+R}\right), \\
\Phi_{4} &= \frac{G m^2 \tau}{\hbar}\left(\frac{1}{d+\sqrt{3}l}+\frac{2}{d+R}\right),
\end{aligned}
\end{equation}
so we can again use $\Delta\Phi_i=\Phi_i-\Phi_1$ to write:
\begin{multline}
    \ket{\psi_{\text{star}}}=\frac{e^{i\Phi_1}}{2\sqrt{2}}\Bigg[\ket{000}+e^{i\Delta\Phi_2}\ket{111}+\\
    e^{i\Delta\Phi_3}\big(\ket{001}+\ket{010}+\ket{100}\big)\\
    +e^{i\Delta\Phi_4}\big(\ket{011}+\ket{101}+\ket{110}\big)\Bigg].
\label{eq:phi_star}
\end{multline}

Defining this time $\mu=e^{i\Delta\Phi_2}$, $\nu=e^{i\Delta\Phi_3}$ and $\xi=e^{i\Delta\Phi_4}$, we can again use (\ref{eq:tangle}) in order to express the three-tangle as:

\begin{multline}
\tau_{\text{star}}(\mu,\nu,\xi)\\
=\frac{1}{16}\Big|\mu^2-3\nu^2\xi^2-6\mu\nu\xi+4\xi^3+4\mu\nu^3\Big|.
\end{multline}

Classifying all the possible states that can be produced in the star setup just by looking at the form of the three-tangle is rather challenging, but it can easily be checked that GHZ states can be created, since, for instance, $\tau_{\text{star}}(1,-1,-1)=\tau_{\text{star}}(-1,1,-1)=1$.

\subsection{General results}

In Figure \ref{fig:neg+wit_vs_time_01}, we show the time evolution of the tripartite negativity and the genuine tripartite entanglement in the three different setups. As we can see, neither of the measures show a strong dependence on the chosen setup, mainly due to the small experimental times that are sensible to consider. In terms of the witness, results show that, for these values of the parameters, the presence of a decoherence of $\gamma=10^{-2}$ Hz is sufficient for genuine entanglement to be undetectable. However, one can also consider different superposition widths while keeping $d_{\text{min}}=35$ $\mu$m fixed. This is displayed in Figure \ref{fig:wit_vs_lgamma_01}, where we show the evolution of the witness with respect to both the decoherence and the superposition width $l$. By inspecting these, we see that choosing $l>10$ $\mu$m can actually lead to better results. With $l=10$ $\mu$m, the witness can withstand up a decoherence of up to $\gamma\simeq0.002$ Hz in all three setups. However, for $l\simeq d_{\text{min}}=35$ $\mu$m, this threshold increases to $\gamma\simeq0.1$ Hz, favouring the detection of entanglement. The results in Figure \ref{fig:wit_vs_lgamma_01} also show that the parameter space is considerably similar for all setups, which allows us to claim that the three of them are nearly equivalent when witnessing genuine tripartite entanglement.

\begin{widetext}

\begin{figure}[t]
    \centering

    \begin{subfigure}{0.49\textwidth}
        \includegraphics[width=
\linewidth]{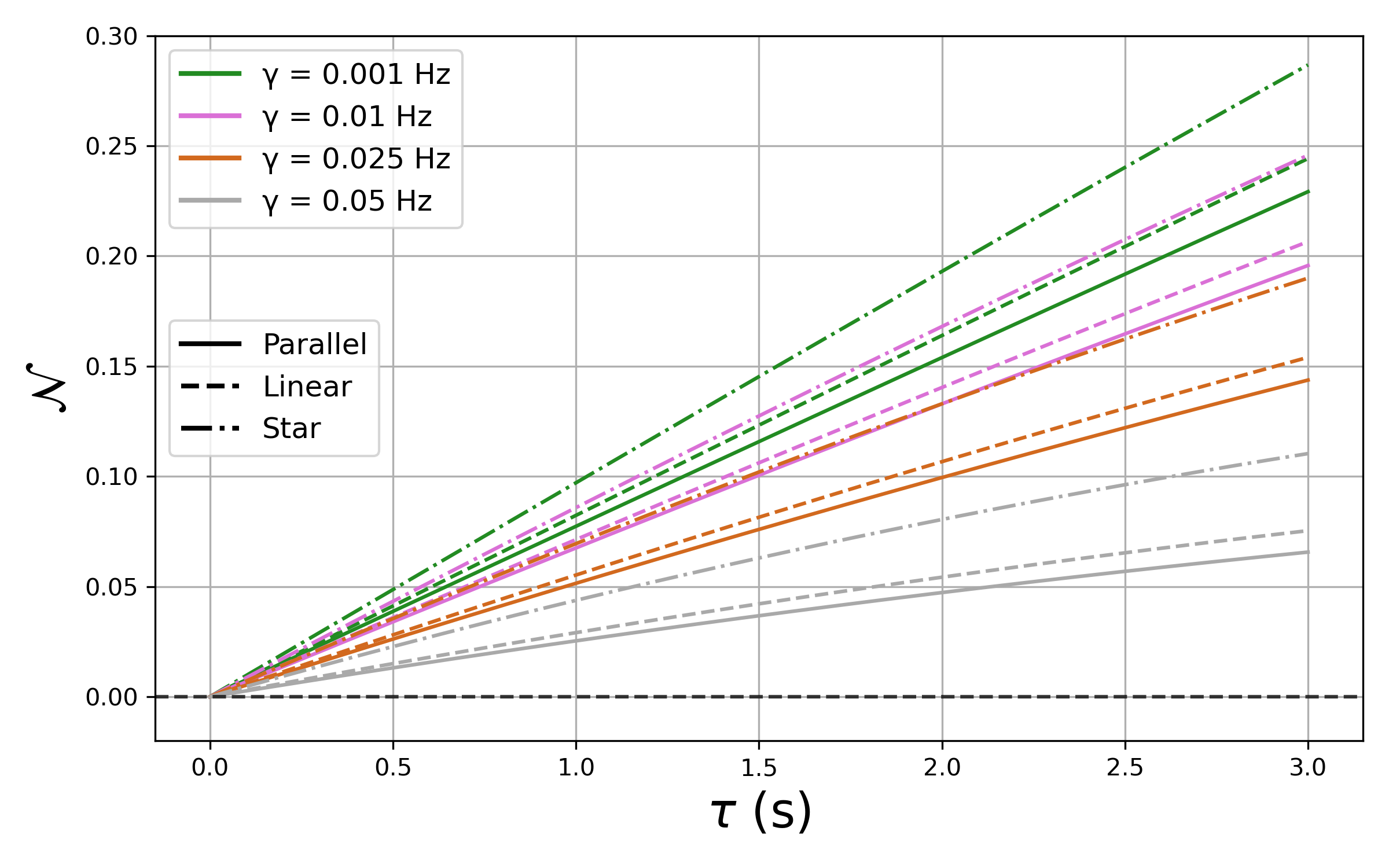}
        \centering
        \caption{}
    \end{subfigure}
    \hfill
    \begin{subfigure}{0.49\textwidth}
    \includegraphics[width=
\linewidth]{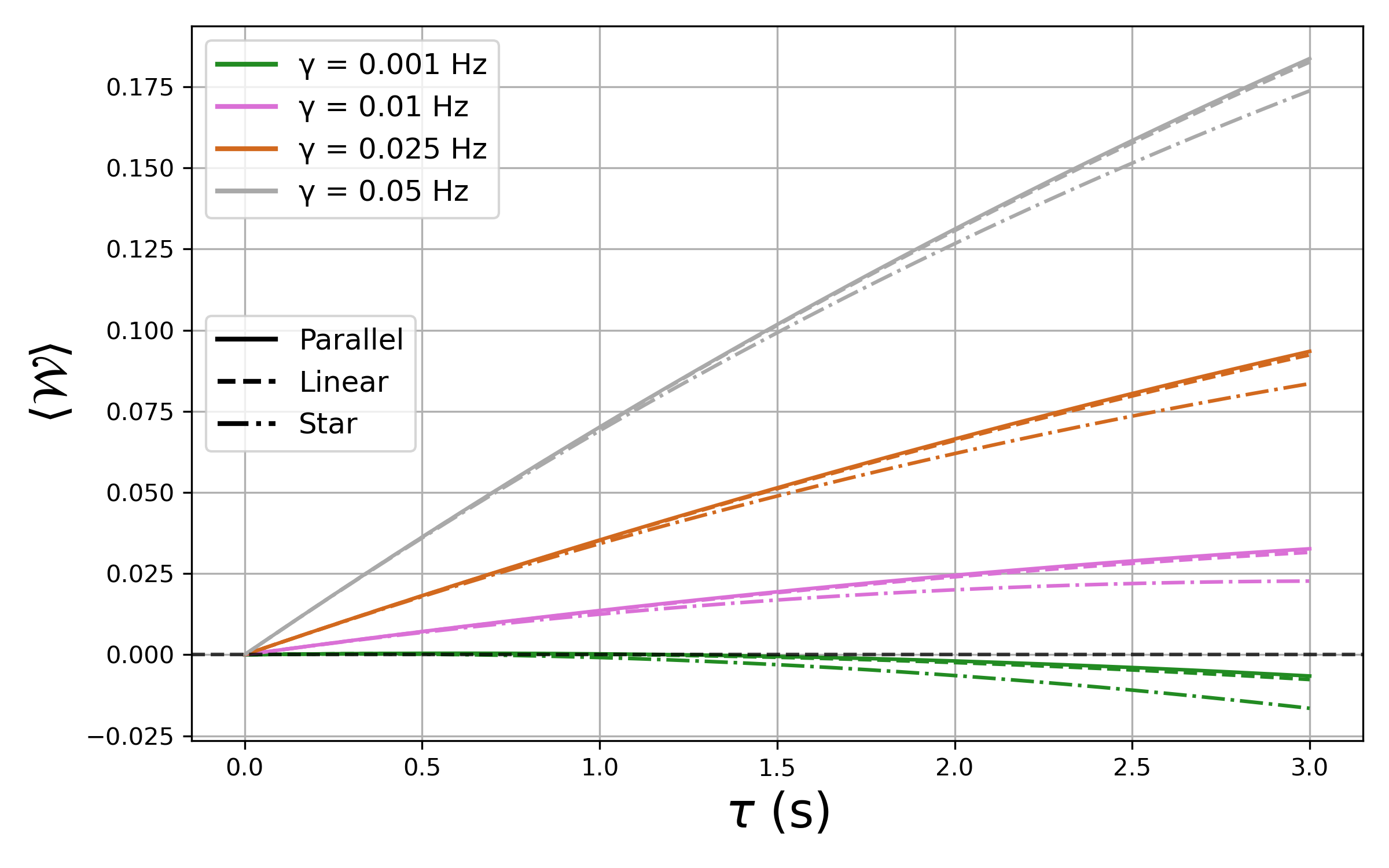}
        \centering

        \caption{}
    \end{subfigure}
    \caption{Time evolution of the (a) tripartite negativity (\ref{eq:trip_neg_def}) and (b) genuine tripartite entanglement witness (\ref{eq:witness_def}) in all three different setups, with different values for the decoherence, as well as the parameters $m=10^{-14}$ kg, $d_{\text{min}}=35$ $\mu$m and $l=10$ $\mu$m, chosen in accordance to \cite{Anupam2}. The tripartite negativity decreases as the decoherence gets larger, while the witness is only capable of detecting genuine tripartite entanglement when $\gamma=10^{-3}$ Hz (out of the values chosen). 
    }
    \label{fig:neg+wit_vs_time_01}
\end{figure}

\begin{figure}[h]
    \centering

    \begin{subfigure}{0.32\textwidth}
        \includegraphics[width=
\linewidth]{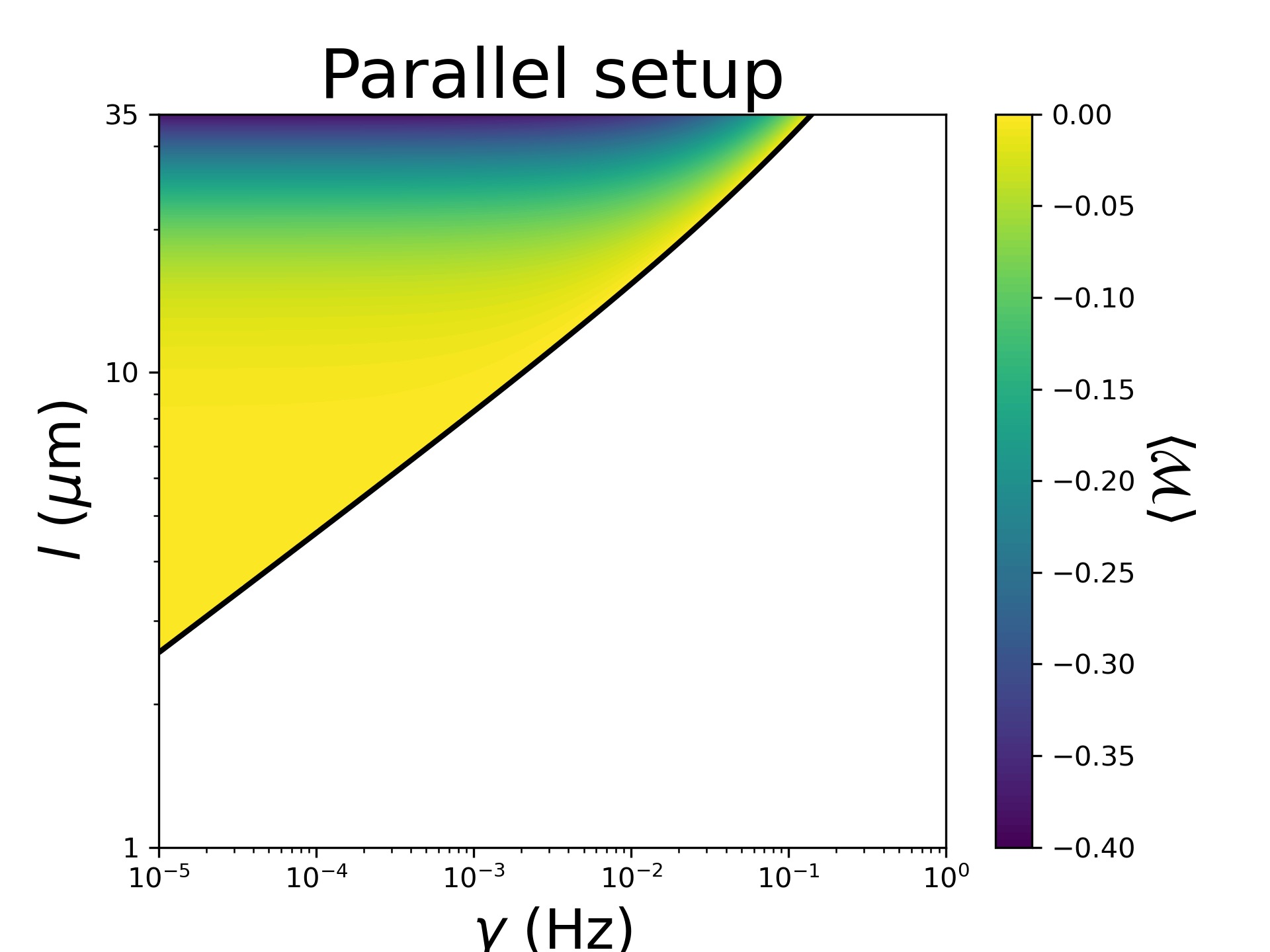}
        \centering
        \caption{}
    \end{subfigure}
    \hfill
    \begin{subfigure}{0.32\textwidth}
    \includegraphics[width=
\linewidth]{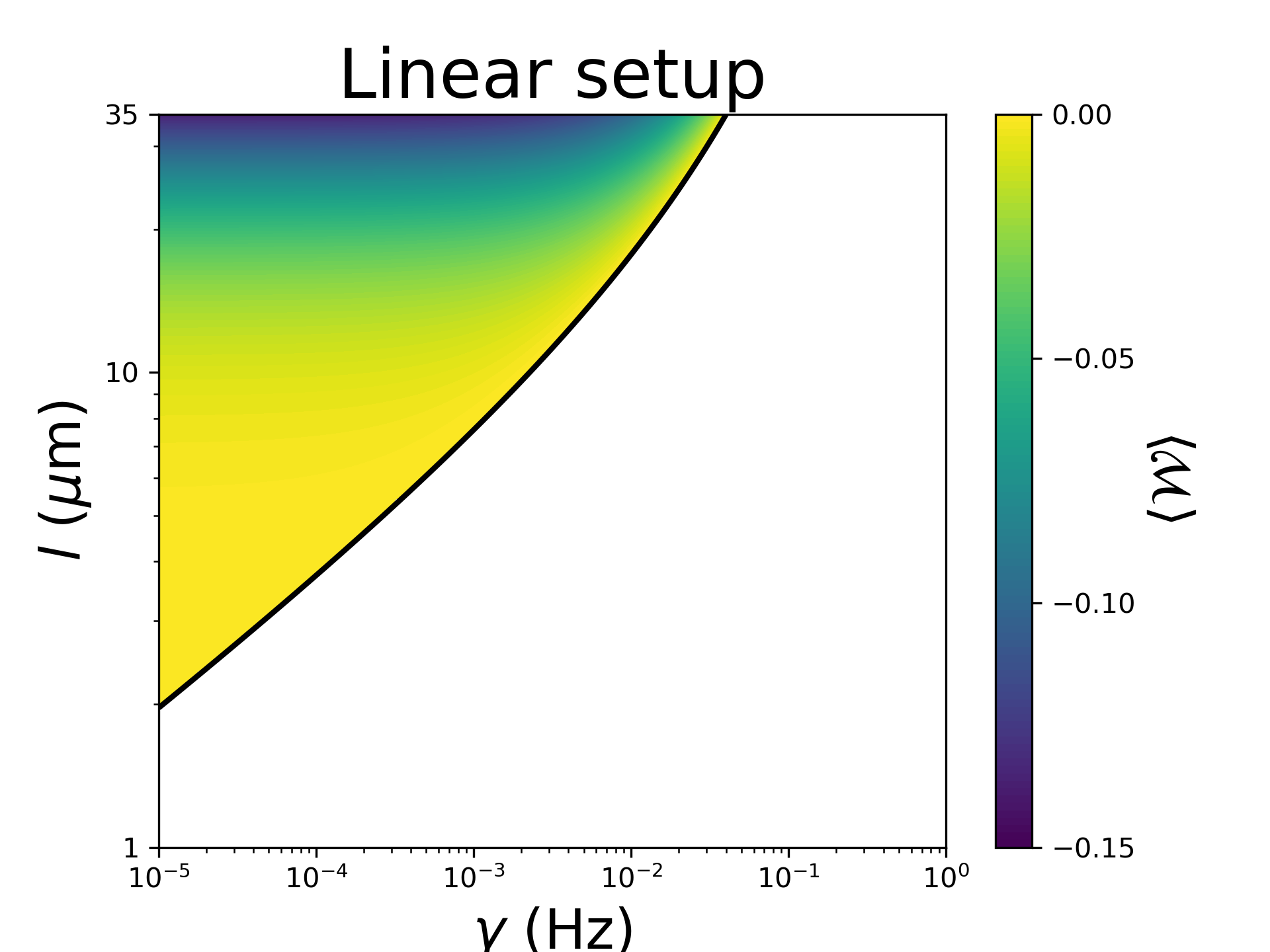}
        \centering
        \caption{}
    \end{subfigure}
    \hfill
    \begin{subfigure}{0.32\textwidth}
    \includegraphics[width=
\linewidth]{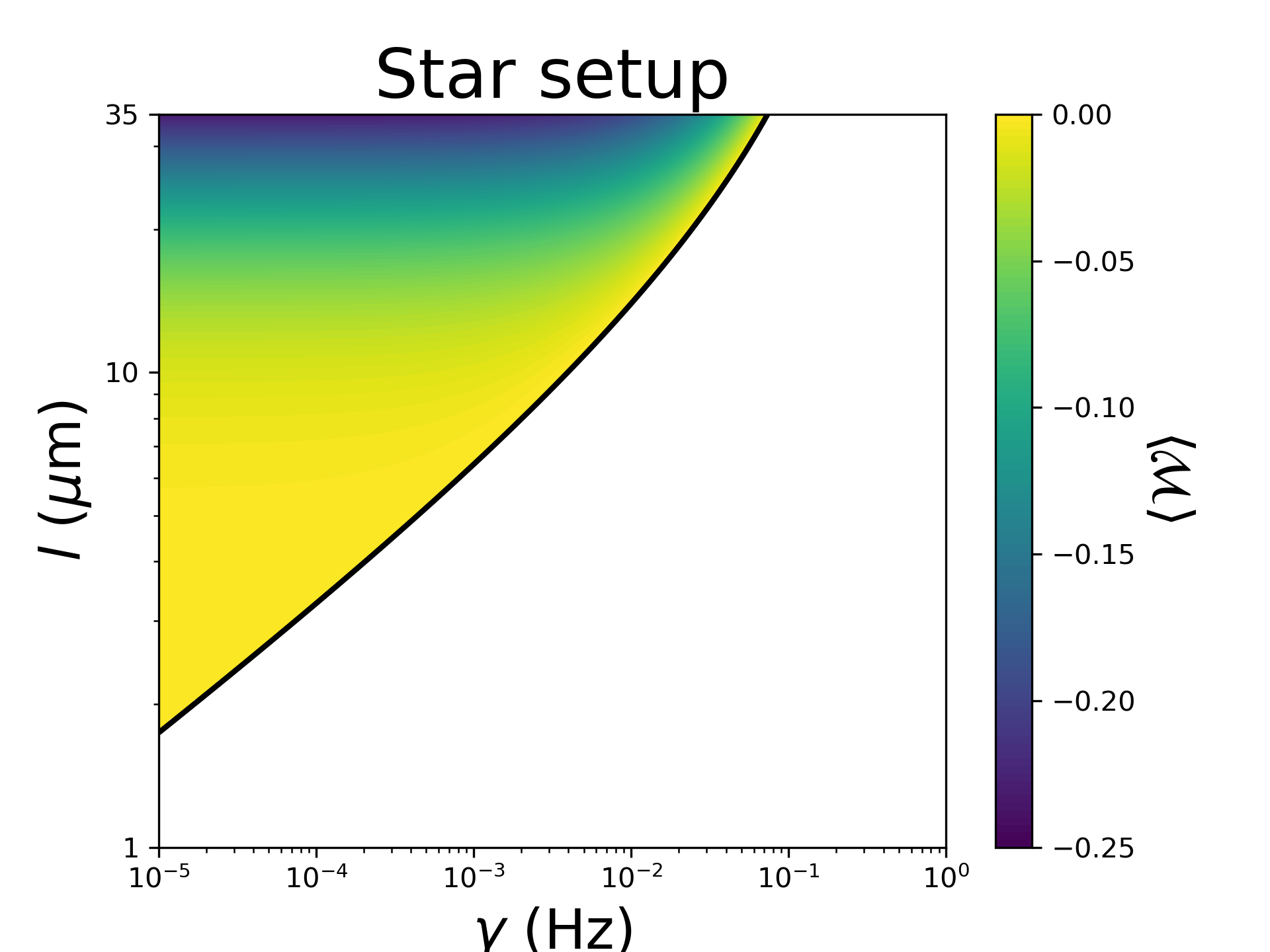}
        \centering
        \caption{}
    \end{subfigure}
    \caption{Genuine tripartite entanglement witness (\ref{eq:witness_def}) values in all three setups as a function of the superposition width $l$ and the decoherence, with $\tau=2.5$ s, $m=10^{-14}$ kg and $d_{\text{min}}=35$ $\mu$m, chosen in accordance to \cite{Anupam2}. Genuine entanglement can be detected with $\gamma\simeq10^{-3}$ Hz for $l=10$ $\mu$m or $\gamma\simeq0.1$ Hz for $l\simeq d_{\text{min}}$.}
    \label{fig:wit_vs_lgamma_01}
\end{figure}

\begin{figure}[h]
    \centering

    \begin{subfigure}{0.32\textwidth}
        \includegraphics[width=
\linewidth]{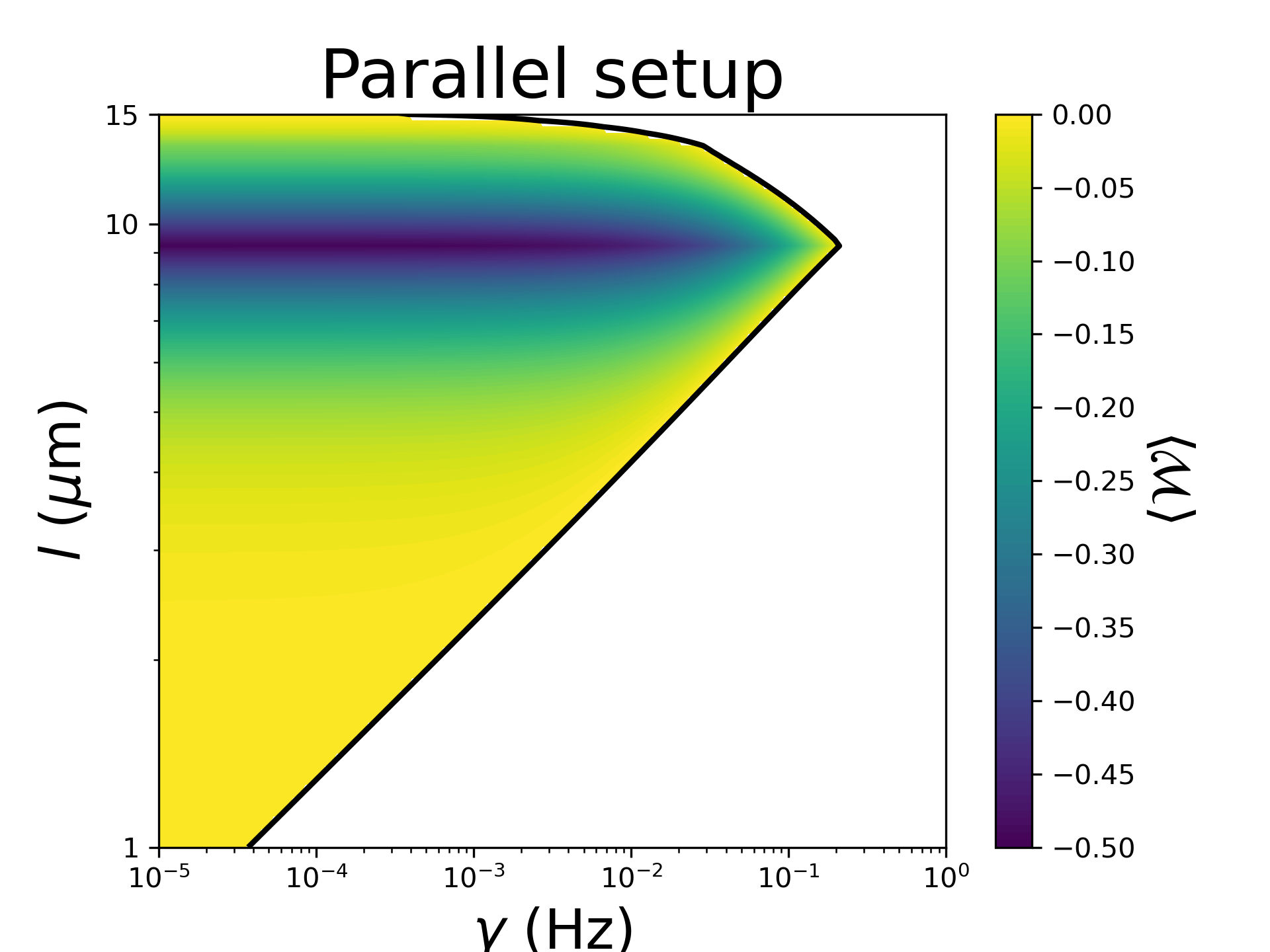}
        \centering
        \caption{}
    \end{subfigure}
    \hfill
    \begin{subfigure}{0.32\textwidth}
    \includegraphics[width=
\linewidth]{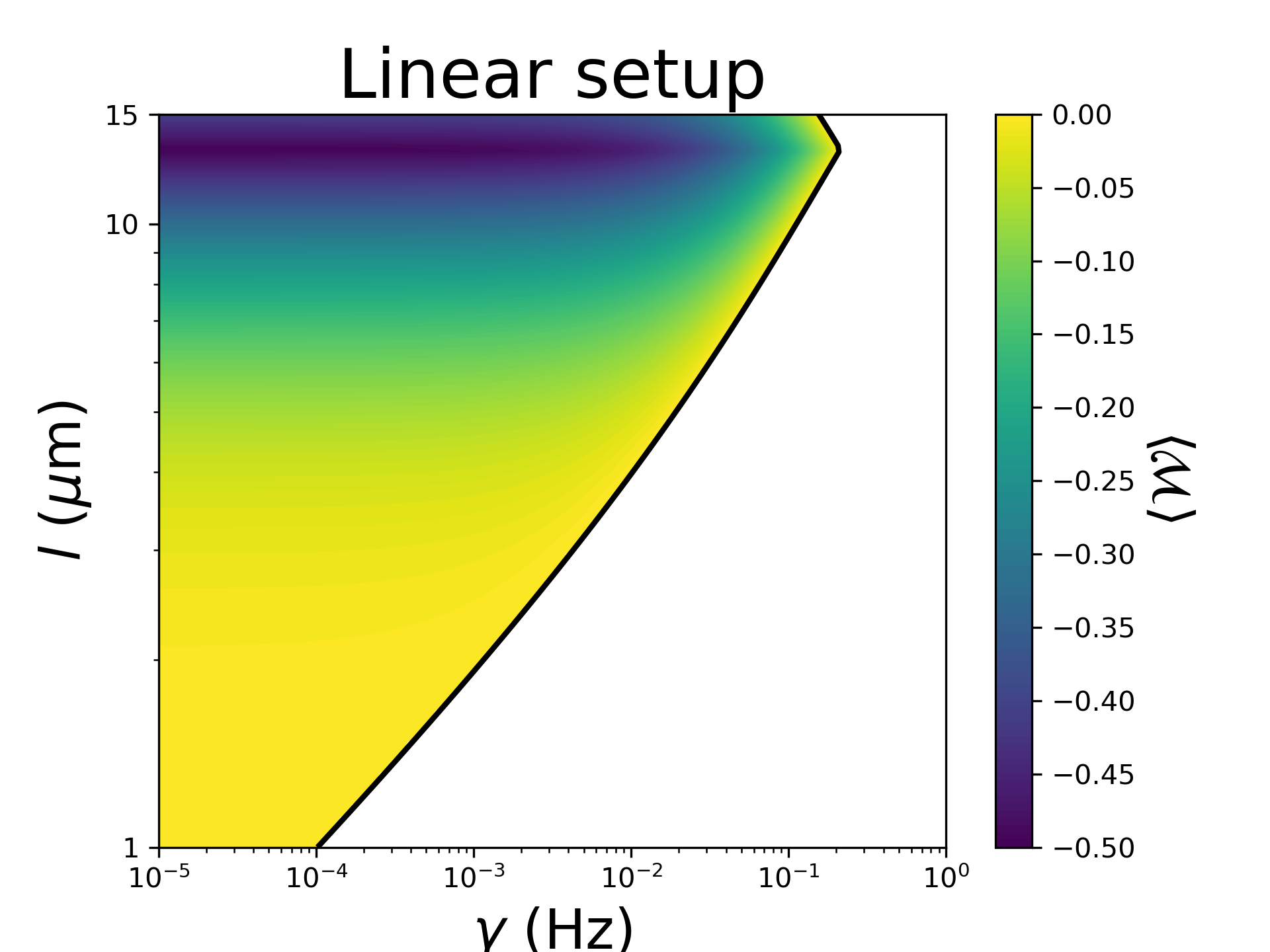}
        \centering
        \caption{}
    \end{subfigure}
    \hfill
    \begin{subfigure}{0.32\textwidth}
    \includegraphics[width=
\linewidth]{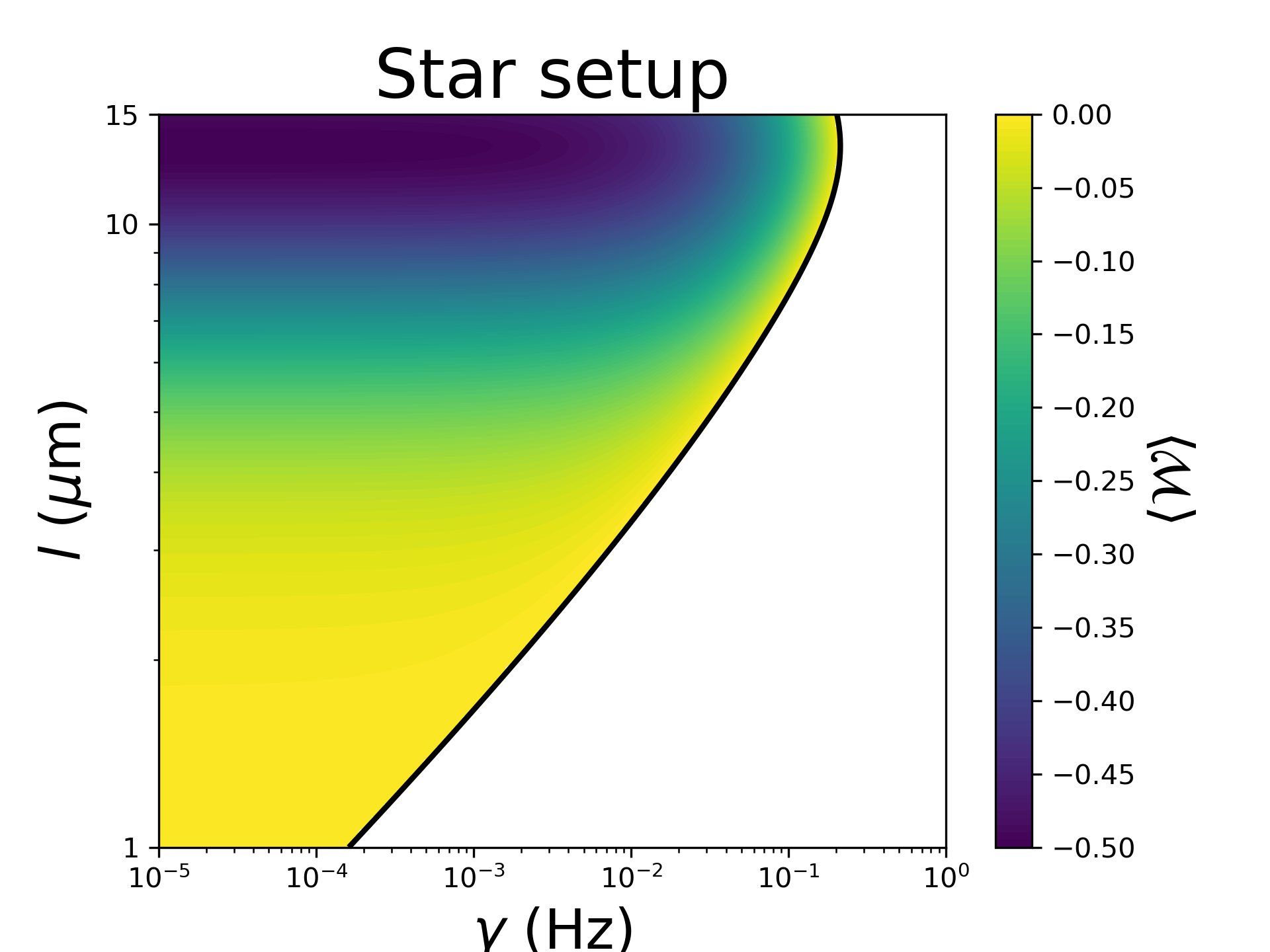}
        \centering
        \caption{}
    \end{subfigure}
    \caption{Genuine tripartite entanglement witness values in all three setups as a function of the superposition width $l$ and the decoherence, with the parameters $\tau=2.5$ s and $m=10^{-14}$ kg, chosen in accordance to \cite{Anupam2}, as well as $d_{\text{min}}=15$ $\mu$m. For this value of the separation, we can find negative values of the witness for $l=10$ $\mu$m and $\gamma \simeq0.1$ Hz, thus making this parameter space more favourable for the detection of genuine tripartite entanglement. }
    \label{fig:wit_vs_lgamma_01_2}
    
\end{figure}

\end{widetext}

There are other changes in the parameters of the QGEM experiment that we may consider. Decreasing the considered value of the mass to $m=10^{-15}$ kg makes the region in which genuine tripartite entanglement is detectable smaller, making it fully impossible with current experimentally feasible values. On the contrary, we noticed that reducing the qubit separation $d_{\text{min}}$ while keeping $l$ constant increases the generated entanglement, as illustrated in Figure \ref{fig:wit_vs_lgamma_01_2}, where we consider $d_{\text{min}}=15$ $\mu$m. We can see that, unlike in the previous case, genuine entanglement can be witnessed with $l=10$ $\mu$m and up to $\gamma \simeq0.1$ Hz. This value for the minimum separation is still relatively far from what is experimentally achievable but it might be considered as a benchmark for future experimental developments.

\section{Conclusions}

In this work, we studied the entanglement that is generated in three-qubit quantum gravity-induced entanglement of masses setups. Our main contributions include the analysis of the evolution of the tripartite negativity and the genuine tripartite entanglement witness for the system in the parallel setup when taking into account the decoherence, which makes the entanglement undetectable when the loss of information of the system to the environment reaches a certain amount that will depend on the values of the physical parameters of the system. We also focused on the importance of calculating a tripartite entanglement measure as opposed to the entanglement of just one bipartition, showing that the existence of individual entanglement does not guarantee genuine entanglement for all values of the parameters in the parallel setup.

Moreover, we made use of another tripartite entanglement measure known as the three-tangle, which helped us check the existing classification of states generated in the parallel setup and extend it to the linear setup, which is a feature that we have not found in the literature. Finally, we presented the time evolution of the system's entanglement in all three setups via the tripartite negativity and a genuine tripartite entanglement witness when a small amount of decoherence is present, showing again the role that the decoherence plays in decreasing the amount of entanglement one can find in the resulting states of the QGEM setups. We proved that, with the current experimental constraints of $d_{\text{min}}=35$ $\mu$m and $m=10^{-14}$ kg, genuine tripartite entanglement can be witnessed in three-qubit QGEM if the decoherence is kept at a maximum of around $\gamma=10^{-3}$ Hz for $l=10$ $\mu$m, or $\gamma\simeq0.1$ Hz for $l\simeq d_{\text{min}}$. Future improvements upon the experimental setups could be able to decrease the minimum separation between any two particles and thus facilitate the detection of genuine tripartite entanglement.


One of the main advantages of the usage of witnesses over other entanglement measures is the ability to implement witnesses experimentally through decompositions into local measurements, which can be done by decomposing the witness into tensor products of Pauli matrices and then optimizing the decomposition in order to minimize the measurement settings \cite{witness3}. This decomposition has been calculated for some specific examples of well-known three-qubit states, as shown in \cite{witness3}, but it is often hard to determine for the arbitrary case \cite{guehne_hannover}. Therefore, in the future, it could be of interest to consider attempting to find the decompositions of the entanglement witness used in this work, which might allow us to perform a statistical analysis similar to the one carried out in \cite{Anupam1,Anupam3} in order to determine the number of measurements one would need to confirm the presence of entanglement in the system with a certain confidence level.



\section*{Acknowledgments}

P.G.C.R. acknowledges:  Grant PRE2022-102488 funded by MCIN/AEI/10.13039/501100011033 and FSE+, with project code: PID2021-127726NB-I00. C.S. acknowledges financial support through the Ramón y Cajal Programme (RYC2019-028014-I) and Consolidación Investigadora (CNS2024-154149). AM's research is funded in part by the Gordon and Betty Moore Foundation through Grant GBMF12328, DOI 10.37807/GBMF12328.

\end{document}